\documentclass{article}
\usepackage{arxiv}

\usepackage[utf8]{inputenc} 
\usepackage[T1]{fontenc}    
\usepackage{hyperref}       
\usepackage{url}            
\usepackage{booktabs}       
\usepackage{amsfonts}       
\usepackage{nicefrac}       
\usepackage{microtype}      
\usepackage{multirow}

\usepackage{amsmath,amsfonts,bm, amsthm}

\def\eqref#1{equation~\ref{#1}}

\def\1{\bm{1}}

\def\rvb{{\mathbf{b}}}

\def\rvg{{\mathbf{g}}}

\def\rvw{{\mathbf{w}}}
\def\rvx{{\mathbf{x}}}
\def\rvy{{\mathbf{y}}}

\def\rmW{{\mathbf{W}}}

\DeclareMathAlphabet{\mathsfit}{\encodingdefault}{\sfdefault}{m}{sl}
\SetMathAlphabet{\mathsfit}{bold}{\encodingdefault}{\sfdefault}{bx}{n}

\newcommand{\R}{\mathbb{R}}

\DeclareMathOperator*{\argmax}{arg\,max}
\DeclareMathOperator*{\argmin}{arg\,min}

\theoremstyle{plain}

\theoremstyle{remark}

\theoremstyle{definition}

\theoremstyle{plain}

\theoremstyle{plain}

\theoremstyle{definition}

\providecommand{\corollaryname}{Corollary}
\providecommand{\lemmaname}{Lemma}
\providecommand{\problemname}{Problem}
\providecommand{\remarkname}{Remark}
\providecommand{\theoremname}{Theorem}

\usepackage{amsmath,epsfig}
\usepackage{amssymb}
\usepackage{amsthm}
\usepackage{algorithm}
\usepackage{algpseudocode}
\usepackage{tikz}
\usepackage{tabu}
\usepackage{graphicx}
\usepackage{subfig}
\usepackage{epstopdf}
\usepackage{epsfig}
\usepackage{booktabs}
\usepackage[flushleft]{threeparttable}
\usetikzlibrary{tikzmark,calc}
\usepackage{float}
\usepackage{multirow}
\usepackage{hyperref}
\usepackage{verbatim}
\usepackage{wrapfig}
\usepackage{appendix}
\usepackage{rotating}
\usepackage{mathtools}
\usepackage{bm}
\usepackage{enumitem}
\usepackage{standalone}
\usepackage{pgfplots}
\usepackage{pgfplotstable}
\usepackage{pifont}

\pgfplotsset{compat=1.14}
\numberwithin{equation}{section} 

\usepackage{microtype}

\def\R{{\mathbb{R}}}

\def\reduce{\textit{reduce}~}

\pgfdeclarelayer{background}
\pgfdeclarelayer{foreground}
\pgfsetlayers{background,main,foreground}

\tikzstyle{sensor}=[draw, fill=blue!20, text width=4em, 
    text centered, minimum height=2.5em]
\tikzstyle{normalCell}=[draw, fill=orange!20, text width=3.2em, 
    text centered, minimum width=2.0em, minimum height=1.4em, 
    rounded corners]
\tikzstyle{conv}=[draw, fill=blue!20, text width=4.2em, 
    text centered, minimum width=2.0em, minimum height=1.4em, rounded corners]
\tikzstyle{operation}=[draw, fill=white!20, text width=3.2em, 
    text centered, minimum width=2.0em, minimum height=1.4em, rounded corners]
\tikzstyle{cell_in}=[draw, fill=green!20, text width=3.0em, 
    text centered, minimum width=2.5em, minimum height=2.3em]
\tikzstyle{cell_out}=[draw, fill=yellow!20, text width=3.0em, 
    text centered, minimum width=2.5em, minimum height=2.3em]
\tikzstyle{square}=[draw, fill=blue!20, minimum size=2em]
\tikzstyle{choice}=[draw, fill=green!20, text width=3.2em, 
    text centered, minimum width=2.0em, minimum height=1.4em, rounded corners]
\tikzstyle{customNode}=[draw, fill=purple!25, text width=3.2em, 
    text centered, minimum width=2.0em, minimum height=1.4em, rounded corners]
\tikzstyle{ann} = [above, text width=5em, text centered]
\tikzstyle{wa} = [sensor, text width=10em, fill=red!20, 
    minimum height=6em, rounded corners, drop shadow]
\tikzstyle{sc} = [sensor, text width=13em, fill=red!20, 
    minimum height=10em, rounded corners, drop shadow]
\tikzstyle{oneShotModel}=[draw, fill=white!10, text width=3.0em, 
    text centered, minimum width=2.0em, minimum height=6.0em]
\tikzstyle{sampleNode}=[draw, fill=white!10, text width=5.0em, 
    text centered, minimum width=2.0em, minimum height=1.0em]
\tikzstyle{CSNode}=[draw, fill=white!10, text width=4.0em, 
    text centered, minimum width=2.0em, minimum height=6.0em]
\tikzstyle{restrictionNode}=[draw, fill=white!10, text width=5.0em, 
    text centered, minimum width=2.0em, minimum height=1.0em]

\usepackage{xintexpr}
\usepackage{siunitx}

\usepackage[compact]{titlesec}
\titlespacing{\section}{0pt}{2ex}{1ex}
\titlespacing{\subsection}{0pt}{1ex}{0ex}
\titlespacing{\subsubsection}{0pt}{0.5ex}{0ex}
\usetikzlibrary{quotes,arrows.meta, shapes.geometric, shapes.misc, fit, positioning, calc, shapes.arrows}

\def\algname{{\textsc{Sphynx}}}
\def\Delphi{{\textsc{Delphi}}}

\title{\algname{}: ReLU-Efficient Network Design \\ for Private Inference}

\usepackage{scrextend}
\deffootnote{1em}{0em}{\thefootnotemark~}

\author{Minsu Cho\textsuperscript{1}, Zahra Ghodsi\textsuperscript{2}, Brandon Reagen\textsuperscript{1}, Siddharth Garg\textsuperscript{1}, and Chinmay Hegde\textsuperscript{1}\\
\textsuperscript{1} New York University, \textsuperscript{2} University of California San Diego \\
\texttt{mc8065@nyu.edu,zghodsi@ucsd.edu,bjr5@nyu.edu,sg175@nyu.edu,chinmayh@nyu.edu} \\
}

\begin{document}

\maketitle

\begin{abstract}
    The emergence of deep learning has been accompanied by privacy concerns surrounding users' data and service providers' models. We focus on private inference (PI), where the goal is to perform inference on a user's data sample using a service provider's model. Existing PI methods for {deep networks} enable cryptographically secure inference with little drop in functionality; however, they incur severe latency costs, primarily caused by non-linear network operations (such as ReLUs). This paper presents \algname{}, a ReLU-efficient network design method based on micro-search strategies for convolutional cell design. \algname{} achieves Pareto dominance over all existing private inference methods on CIFAR-100. We also design large-scale networks that support cryptographically private inference on Tiny-ImageNet and ImageNet. 
\end{abstract}

\section{Introduction}
\label{sec: introduction}

Deep {learning} inference is often outsourced to external cloud services to mitigate the high cost of executing state-of-the-art deep networks on user devices~\cite{li2020automating}. However, outsourced inference raises essential concerns about data and model privacy: users may not trust a cloud service provider with their data, and cloud service providers may not want to share their models, trained at enormous expense, with users.
Private inference (PI) provides a solution to this problem by guaranteeing user and model privacy using cryptographic techniques~\cite{juvekar2018gazelle, DELPHI}. Under PI, a user can perform inference using a model hosted in the cloud without the cloud learning anything about her data and, conversely, without the user learning anything about the cloud's model parameters. 

PI techniques thus far reported in the literature have leveraged a range of cryptographic protocols, including homomorphic encryption (HE), additive secret sharing (SS), and garbled circuits (GC). However, these all incur heavy computational overheads, resulting in several orders-of-magnitude increase in inference latency compared to standard ``plaintext'' inference. {Prior work has demonstrated} that \emph{non-linear} network operations like the Rectified Linear Unit (ReLU) and max-pooling are the key bottlenecks. For example, Ghodsi et al.~\cite{NEURIPS2020_c519d47c} estimate that ReLU layers in {MiniONN~\cite{liu2017oblivious} are four orders of magnitude more expensive than convolution layers, while  
in \Delphi{}~\cite{DELPHI}, 
ReLUs account for $93\%$ of ResNet32's online private runtime~\cite{DEEPREDUCE}.}
This is in direct contrast to standard (plaintext) inference where ReLUs and max-pools are effectively free and the dominant runtime costs are due to the floating-point operations (FLOPs) of convolutional and fully-connected layers. 



With this in mind, recent efforts have sought to re-think deep neural network architectures that mitigate the dominant costs of ReLUs in PI run-time. Methods like \Delphi~\cite{DELPHI} and DeepReDuce~\cite{DEEPREDUCE} start with state-of-the-art architectures for standard plaintext inference (such as VGG or ResNets), and replace carefully selected ReLUs with quadratic activations (in the case of \Delphi) or with the identity function (in the case of DeepReDuce).
{SAFENet~\cite{SAFENET} extends \Delphi{}, substituting ReLUs channel-wise with polynomial activations of variable degree.}

A parallel line of research, pioneered by CryptoNAS~\cite{NEURIPS2020_c519d47c}, is to \emph{re-design} neural architectures from scratch for ``PI-efficiency.''
This can be accomplished using neural architecture search (NAS) techniques with the objective of minimizing ReLU operations in place of traditional FLOP count optimization.
Specifically, CryptoNAS executes a modified version of \emph{macro-search}~\cite{zoph2016neural}, i.e., over entire convolutional {neural network models} with arbitrarily many skip connections between layers. However, a long line of standard NAS research~\cite{pham2018efficient, liu2018darts, Xu2020PC-DARTS:, xie2018snas, dong2019search} has shown that searching over a {smaller} \emph{micro-search} space of convolutional cells, which are then repeated across layers, yields more accurate networks for standard (plaintext) inference. In this paper we examine if the same is true in the case of private inference: can we outperform CryptoNAS (and other state-of-the-art PI methods including \Delphi, SAFENet, and DeepReDuce) via micro-search principles? 
To this end, this paper proposes \algname{}, a new framework for ReLU-efficient micro-search to design convolutional cells that maximize accuracy under a constraint on the number of ReLUs. 

\algname{} enables cell-based NAS algorithms (such as DARTS~\cite{liu2018darts}) to discover novel \textit{normal} and \textit{reduce} cells for highly-accurate models that use significantly fewer ReLU/max pool operations.
Moreover, unlike conventional micro-search techniques (which follow the search space introduced in NASNets~\cite{zoph2018learning}), \algname{} also learns the 
{best}
location of \textit{reduce} cells, which turns out to be critical for minimizing ReLU costs. We perform extensive ablation studies to analyze the utility of each component of \algname{} in detail. Overall, the innovations of \algname{} are threefold:
\begin{enumerate}[leftmargin=*,nosep]
    \item We propose a new ReLU-efficient micro-search space for cell-based NAS. This new search space can be used in conjunction with any of several existing search strategies (such as DARTS, ENAS~\cite{pham2018efficient}, PC-DARTS~\cite{Xu2020PC-DARTS:},  GDAS~\cite{dong2019search}, and GAEA~\cite{li2021geometryaware}) to discover efficient \textit{normal}/\textit{reduce} convolutional cells. 
    \item 
    We show that unlike traditional micro-NAS,
    where \reduce cells are uniformly spaced,
    the precise location of \reduce cells is critical in maximizing accuracy under ReLU constraints. 
    To optimally position the \reduce cells we develop a novel post-processing method based on the Gumbel-Softmax re-parameterization trick that helps further improve network performance. 
    \item We {demonstrate the effectiveness of \algname{} using}
    image classification datasets including CIFAR-100~\cite{CIFAR100}], Tiny-ImageNet~\cite{Le2015TinyIV}, and ImageNet~\cite{ILSVRC15} and 
    demonstrate that \algname{} outperforms several state-of-the-art solutions for PI. 
     
\end{enumerate}

\section{Background on Private Inference}
\label{sec: pibackground}

\algname{} uses the \Delphi{} protocol~\cite{mishra2020delphi} for PI under the same threat model and provides identical security guarantees. Specifically, \Delphi{} models two parties, a server that holds a trained model for an $L$ layer deep network with parameters $\rmW_i$ and $\rvb_i$ for layer $i$ ($i\in[0,L-1]$), and a client holds that holds input $\rvy_0$. 
The client seeks to obtain $\rvy_{L}$, computed layer-wise as follows: 
$\rvy_{i+1} = \sigma_i \left( \rmW_i \rvy_i + b_i \right)$, where $\sigma_i$ is a non-linear activation function,
ReLU is used here.

The parties are assumed to be semi-honest, i.e., they follow the protocol faithfully but try to infer information about the other party's input. That is, the client seeks to recover $\rmW_i$ and $\rvb_i$ and the server seeks to recover $\rvy_0$. The goal of PI is to prevent the parties from learning \emph{anything} about the other parties data (other than what the client can learn from output $\rvy_{L}$). 

\Delphi{} builds on three cryptographic primitives: additive secret sharing (SS) for linear layers, garbled circuits (GC) for non-linear ReLU layers, and homomorphic encryption (HE) used offline. We begin by briefly reviewing these primitives.

\textbf{Additive Secret Sharing}~\cite{shamir1979share} allows two parties to hold \emph{additive} shares $[x]_1$, $[x]_2$ of a secret value $x$ (defined over a field $F_p$) such that $[x]_1+[x]_2 = x$. The shares can be generated by sampling a random value $r$ and setting $[x]_1 = r$ and $[x]_2 = x-r$.

\textbf{Garbled Circuits}~\cite{yao1986generate} is a scheme introduced by Yao that allows two parties to compute a Boolean function $f$ on their private inputs without revealing their inputs to each other. The function $f$ is first represented as a Boolean circuit of two-input logic gates. 
One of the parties, the garbler, encodes (garbles) the circuit by encrypting the truth table of each gate in the circuit and sends the resulting ``garbled circuit" to the other party, the evaluator. 
The other party, the evaluator, computes (or decrypts) the circuit gate-by-gate using encodings of the garbler's inputs and her own inputs, producing an encoding of the circuit's output.
She shares this encoding with the garbler, who then reveals the corresponding plaintext.  

\textbf{Homomorphic Encryption}~\cite{gentry2011implementing} enables operations on encrypted values without a private key or decryption. A cryptosystem supports a homomorphic operation ($*$) if for public key ($pk$), secret key ($sk$), and ciphertexts $c_1=\text{Enc}(pk, m_1)$, $c_2=\text{Enc}(pk, m_2)$, there exists a function $\textsc{Eval}$ such that $\text{Dec}(sk, \textsc{Eval}(c_1, c_2)) = m_1 * m_2$. A fully homomorphic cryptosystem supports arbitrary homomorphic additions and multiplications.

\Delphi{} is a hybrid protocol that uses different cryptographic primitives for linear and ReLU layers. 
Specifically, \Delphi{} uses HE ({offline}) and SS ({online}) for linear layers, and GC for ReLUs. We summarize these protocols below, assuming $\rvb_i = 0$ for simplicity.   

\emph{Linear layers:} During the offline phase, the client and server sample random vectors $\mathbf{r_i}$ and $\mathbf{s_i}$ respectively for linear layer $i$. The client encrypts and sends $Enc(pk, \mathbf{r_i})$ to the server. The server computes $Enc(\mathbf{W_i}.\mathbf{r_i}-\mathbf{s_i})$ homomorphically, and sends it back to the client. The client decrypts this ciphertext, and obtains $\mathbf{W_i}.\mathbf{r_i}-\mathbf{s_i}$,
which will be used later in the online phase. 

During the online phase, the client computes and sends $\mathbf{y}_{0}-\mathbf{r_0}$ 
to the server. The client and server now hold secret shares of the client's input $\mathbf{y}_{0}$, or equivalently, the first layer's inputs.  
The server then computes $\mathbf{W_0}.(\mathbf{y_0}-\mathbf{r_0})+\mathbf{s_0}$, its own share of the first layer's output.
Correspondingly, the client's share of the first layer's output is $\mathbf{W_0}.\mathbf{r_0}-\mathbf{s_0}$, which was already obtained in the offline phase. It can be confirmed that the shares sum to $\mathbf{W_0}.\mathbf{y_0}$, i.e., the client and server each hold an additive secret share of $\mathbf{W_0}.\mathbf{y_0}$. The same protocol is used in subsequent layers allowing the client and server to obtain shares of each linear layer's outputs from shares of its inputs. 
Importantly, the only computation performed online is the server's computation of its share, which can be performed at roughly the same speed as the layer's plaintext computations~\cite{DELPHI}.   

\emph{ReLU layers:} During the offline phase, the server garbles the circuit representing the ReLU function and sends it to the client, along with encodings of $\mathbf{r_{i+1}}$ and $\mathbf{W_i}.\mathbf{r_i}-\mathbf{s_i}$. During the online phase, the server sends the encodings of $\mathbf{W_i}.(\mathbf{y_i}-\mathbf{r_i})+\mathbf{s_i}$ to the client, who is now able to evaluate the garbled circuit and send the encoded output to the server.
The server decodes the garbled circuit output, which is $\mathbf{y_{i+1}}-\mathbf{r_{i+1}}$, the server's share for the next linear layer.

From the above discussion, we can observe that linear layers' online computations are effectively the same as plaintext computations and can be accelerated using standard CPU/GPU libraries. On the other hand, operations involving ReLU (or max-pool) layers use GCs that require expensive online crpytographic computations and interaction between parties, resulting in high latency. This motivates the need for a systematic method to design PI-efficient neural architectures that judiciously minimize ReLU computations, which we address next.

\section{Limitations of Existing NAS Micro-Search Methods for Private Inference}

Before introducing \algname{}, we first analyze conventional NAS micro-search methods \cite{zoph2018learning, liu2018darts, Xu2020PC-DARTS:, dong2019search, xie2018snas, li2019random, cho2019one} through the lens of private inference.
{We conclude that networks found using existing micro-cell search are ill-suited for PI, motivating the need for \algname{}}.
This is because existing architectures found via NAS micro-search use far too many ReLU operations, making such networks prohibitively slow for private inference. 




We start with a NAS primer. 
Typical NAS micro-search techniques seek two types of \emph{cells}: \textit{normal} and \textit{reduce} cells. 
Cells are essentially small networks of layers with convolutions, ReLUs, pools, batch norms (BN), and skip connections. \textit{Normal} cells learn finer-scale features and their outputs have the same spatial resolution as their inputs. \textit{Reduce} cells decrease spatial resolution (usually with stride 2). 

Both types of cells are defined using a directed acyclic graph representation consisting of $N$ nodes.
Each node $x^{(i)}$ represents a feature map, and each directed edge $(i, j)$ is associated with some operation $o^{(i, j)}$ that transforms $x^{(i)}$. For example, the cell representation used in DARTS~\cite{liu2018darts} has $N{=}7$ nodes --- two input nodes, one output node and four intermediate nodes --- with eight operation edges. The output node concatenates features from all intermediate nodes. Each intermediate node is computed by summing all of its parent nodes: $x^{(j)} = \sum_{i<j} o^{(i, j)}(x^{(i)})$. 
For concreteness, let us focus on the micro-search space used in DARTS~\cite{liu2018darts}.
Consider, first, the operation set $\mathcal{O}$. Conventionally, $\mathcal{O}$ is described as a set containing eight different operations: 3\texttimes3 and 5\texttimes5 separable or dilated separable convolutions, 
3\texttimes3 max pooling, 3\texttimes3 average pooling, identity, and \textit{zero}. Here, each ``convolution'' operation is actually a sequence of ReLU-Conv-BN operations. 
Therefore, a feature map size $\{H_i, W_i, C_i\}$ in the $i^{\text{th}}$ cell implies that each convolution operation computes $H_i{\times} W_i{\times} C_i$ ReLU operations. 


\textcolor{black}{However, additional costs emerge. Each cell takes two previous cells' output as inputs. Recalling that the cell output channel dimension is four times larger than its input, the output channel number requires a dimensionality reduction equivalent to the input dimension. To reconcile this difference, a particular \emph{pre-processing} layer, a sequence of ReLU-Conv1\texttimes1-BN, is inserted to downscale the number of channels back to the desired resolution. For the $i\textsuperscript{th}$ cell, this imposes additional $2 {\times}4{\times}H_i {\times} W_i {\times} C_i$ ReLU operations (2 and 4 comes from number of inputs and filters, respectively).}
 \begin{table}[ht]
 \centering
    \resizebox{0.42\textwidth}{!}{
    \begin{tabular}{|c | c | c |}
    \hline
    \multirow{2}{*}{\textbf{Network Setup}} & \multicolumn{2}{c|}{\textbf{ReLU Counts}} \\
    \cline{2-3} & DARTS & PC-DARTS \\
    \hline
    C=36, D=20 & 7796K & 8073K  \\ 
    C=5, D=10  & 523.5K & 591.4K \\
    C=5, D=5 & 231.6K & 314.9K \\
    C=2, D=4 & 73.2K & 107.5K \\
    C=1, D=4 & 36.6K & 53.8K \\
    \hline
    \end{tabular}}
    \vspace{1em}
    \caption{\sl  ReLU count comparisons. `C' and `D' stand for initial channels and network depth, respectively; $C{=}36$ and $D{=}20$ are default hyperparameters to get good performance on CIFAR-10 with DARTS and PC-DARTS.} 
    \label{table: ReLU Counts}
 \end{table}

In this manner, we see that the DARTS search space imposes severe costs when viewed from the perspective of ReLU counts (a fact that seems to have been overlooked by traditional NAS methods, which optimize for FLOP counts). If we want to design a network satisfying a ReLU budget of 90K (which roughly translates to PI latency of 2 seconds using the \Delphi{} protocol) using cells reported by DARTS. 
Using even a very small network with an initial number of $C{=}5$ channels, even the first \textit{normal} cell from this small network contains roughly 95K ReLU operations. Table~\ref{table: ReLU Counts} compares ReLU counts of cells reported by DARTS and PC-DARTS~\cite{Xu2020PC-DARTS:}; 
note that even small networks of depth $D{=}4$ and $C{=}2$ with PC-DARTS cells require more than 100K ReLU operations. 

\begin{figure}[!t]
    \centering
    \begin{tabular}{c}
    \hspace{-6.0em}
    \input{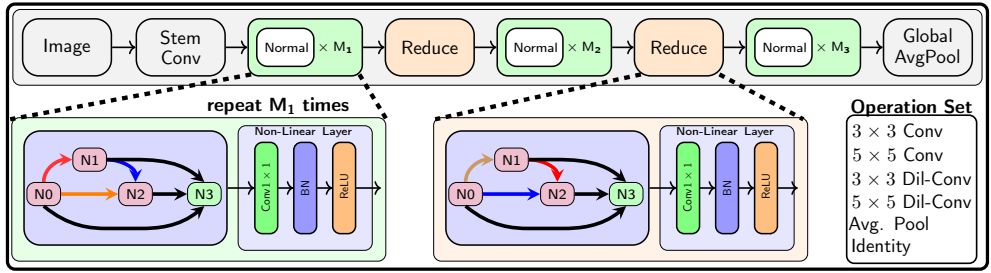}
    \end{tabular}
    \vspace{0.5em}
    \caption{\sl\algname{} search space network skeleton contains \textit{normal} and \textit{reduce} cells. The numbers of $M_1$, $M_2$, and $M_3$ represent the number of repeating \textit{normal} cells, depending on the location of reduce cells. Note that $M_1{=}M_2{=}M_3$ in conventional micro-search NAS literature. For example, consider the final architecture with 8 cells and the possible reduce location is $\{0,1,2, \ldots, 7\}$. If the reduce cells are located at 1 and 3, $M_1{=}1$, $M_2{=}1$, and $M_3{=}4$. Note that \textit{normal} and \textit{reduce} cells do not contain ReLU operations, and 1\texttimes1 Conv-BN-ReLU modules follow both \textit{normal}/\textit{reduce} cell's end. Operation set does not include max-pooling operation due to its non-linearity.}
    \label{fig: search space}
\end{figure}

\section{The \algname{} Framework}
\label{sec: proposed algorithm}

We now propose \algname{}, a new network design approach for efficient private inference. \algname{} consists of three components: (i) a new ReLU-efficient search space, which can be combined with any existing search methods to discover promising \textit{normal} and \textit{reduce} cell architectures; (ii) an approach to select the initial number of channels ($C$) and network depth ($D$) in order to satisfy a ReLU budget; and (iii) a new stochastic optimization method to optimally discover the locations of \textit{reduce} cells in terms of layer depth.
We note that conventional NAS approaches also adopt (ii) (albeit while optimizing for FLOP budgets). To the best of our knowledge, (iii) is novel and could be of independent interest in NAS research.

\subsection{\algname{}: Search Space}
\label{subsec: Efficient search space}
\begin{figure}[!t]
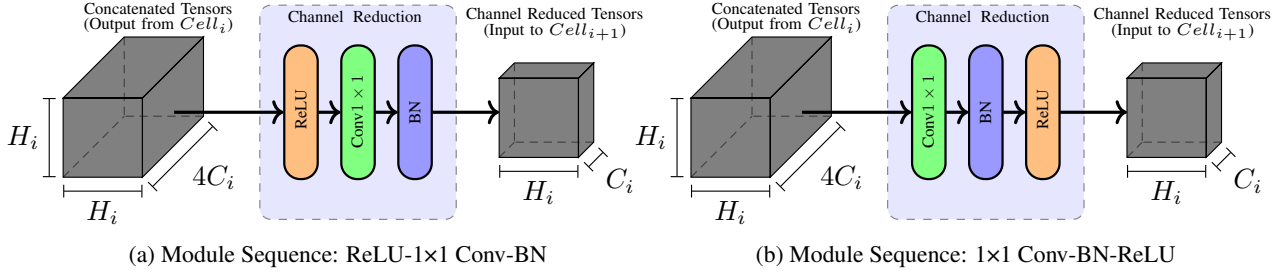

    \centering
    \begin{tabular}{c c}
        \resizebox{0.48\textwidth}{!}{\input{tikz/relu_loc}} & 
        \resizebox{0.48\textwidth}{!}{\input{tikz/relu_loc2}} \\
        \multicolumn{1}{c}{\small (a) Module Sequence: ReLU-1\texttimes1 Conv-BN} & 
        \multicolumn{1}{c}{\small (b) Module Sequence: 1\texttimes1 Conv-BN-ReLU} \\
    \end{tabular}
    \vspace{1.5em}
    \caption{\sl Visualization of ReLU counts on two different module sequence. (a) The original convolution module sequence requires the ReLU count $4C_i{\times}H_i{\times}H_i$. (b) Our proposed search space applies ReLU layers, saving ReLU counts by a factor of four.}
    \vspace{0.7em}
    \label{fig: relu layer visualization.}
\end{figure}

We first describe a new ReLU-efficient search space, where we redefine the skeleton of the network; see Figure~\ref{fig: search space}. Compared to the traditional (DARTS) search space, we propose the following changes: 

\begin{enumerate}[leftmargin=*,noitemsep]
    \item We eliminate the ReLU layer from convolution operations so each ReLU-Conv-BN sequence is now simply a Conv-BN sequence. We also replace separable convolutions with vanilla convolutions. Vanilla convolutions are more expressive than separable convolutions, at the expense of more FLOPs.\footnote{In private inference, FLOP count is not a concern compared to ReLU counts.}
    \item We remove all max-pooling operations since these also require expensive GCs to compute. However, we retain average-pooling operations since they are linear and can be efficiently computed in \Delphi. 
    \item Instead of reducing the dimensionality of a cell's inputs via pre-processing, we instead reduce the dimensionality of a cell's outputs in a single \emph{post-processing} step. Further, as shown in Figure~\ref{fig: relu layer visualization.}, post-processing is performed using a Conv1\texttimes1-BN-ReLU sequence instead of ReLU-Conv1\texttimes1-BN so that the ReLU operate on a tensor with $4\times$ few channels. The cell's outputs are forwarded the subsequent two cells.
\end{enumerate}

Steps 1 and 2 remove all non-linear operations from the DARTS search space. The only non-linearity in our new \algname{} search space is at the output of each cell, as described in Step 3.

In addition, \algname{} follows the ReLU balancing rule introduced in CryptoNAS~\cite{NEURIPS2020_c519d47c}. Unlike conventional FLOP balancing methods which doubles the channel size when the spatial resolution is halved, ReLU balancing quadruples channel size in order to distribute ReLU equally across layers. This strategy has been shown empirically to achieve higher accuracy for low ReLU budgets~\cite{NEURIPS2020_c519d47c}.

\subsection{\algname{}: Search phase}
\label{subsec: searching cells to sphynx space}

\noindent\textbf{Finding cells.} Having defined the search space, we now discover \textit{normal}/\textit{reduce} cells using the DARTS micro-cell search algorithm; for complete details refer Appendix~\ref{appendix: darts background}. The choice of DARTS here is entirely for convenience, and \algname{} can be used in conjunction with any other search method such as ENAS~\cite{pham2018efficient}, GDAS~\cite{dong2019search}, PC-DARTS~\cite{Xu2020PC-DARTS:}, and GAEA~\cite{li2021geometryaware}.



\noindent\textbf{Choosing initial number of channels and network depth.} Once we have discovered \textit{normal}/\textit{reduce} cells, we need to determine the initial channels ($C$) and depths ($D$) of the network to satisfy the ReLU budget constraint. Intuitively, the total number of ReLU operations scales with both $C$ and $D$. Furthermore, due to the ReLU balancing rule, the ReLU counts on each \emph{post-processing} layer are equally distributed. Therefore, the total number of ReLU in the network is simply $H_0{\times}W_0{\times}C{\times}D$, where $H_0$ and $W_0$ are height and width of initial feature map, respectively. \textcolor{black}{We empirically observed that the network performance deviates very little from $C$ and $D$ selections given a ReLU budget. We support our observation with results from ablation experiments conducted with various $C$ and $D$ selections given 50K ReLU budget in Appendix~\ref{appendix: additional ablation studies}}.

  
    \begin{algorithm}[!t]
    \caption{Pseudocode: Finding location of reduce cells}
    \begin{algorithmic}[1]
        \State \textbf{Inputs: } Split train dataset to $D_T$ and $D_V$. 
        \While {not $\hat{\bm{\beta}}$ not converged} 
            \State Sample the minibatch $(\rvx_T, \rvy_T)$ from $D_T$.
            \State Sample the candidate network from $\bm{\beta}$ using Eq.~\ref{eq: gumble max trick}.
            \State Calculate $\sum_{(x, y) \in (\rvx_T, \rvy_T)}\mathcal{L}(F_{\rvw^*, \bm{\beta}}(x), y)$ and update $\rvw$.
            \State Sample the minibatch $(\rvx_V, \rvy_V)$ from $D_V$.
            \State Sample the candidate network from $\bm{\beta}$ using Eq.~\ref{eq: gumble max trick}.
            \State Calculate $\sum_{(x, y) \in (\rvx_V, \rvy_V)}\mathcal{L}(F_{\rvw^*, \bm{\beta}}(x), y)$ and update $\bm{\beta}$ via Eq.~\ref{eq: gumbel softmax approximation}
        \EndWhile
        \State Pick the architecture candidate via Eq.\ref{eq: pick candidate}
    \end{algorithmic}
    \label{alg: reduce cells loc}
    \end{algorithm}

\noindent\textbf{Choosing the location of reduce cells.} Conventional cell-based NAS approaches, including DARTS, fix the position of \textit{reduce} cells at $D/3$ and $2D/3$ cell-depth index. In contrast, we propose a method to find the optimal position of \reduce cells to improve network performance\footnote{In practice, due to channel scaling effects we find that re-locating \textit{reduce} cells increases overall parameter count, but again this is not a bottleneck for private inference since we are focused on optimizing for ReLUs.}.

We focus on the case with two \reduce cells; extending this to multiple cells is straightforward. Given a network with $D$ cells, let $\bm{\beta} \in \R^K$ be a position indicator vector where $K{=}\binom{D}{2}$ is number of all possible candidates of \reduce cell locations, and let $\hat{\bm{\beta}} = \text{softmax}(\bm{\beta}) = \frac{\exp{{\beta}}_i}{\sum_{k} \exp{{\beta}}_k}$ be a probability distribution over $K$ elements. 

\begin{figure}[!t]
    \centering
    \resizebox{0.75\textwidth}{!}{
    \input{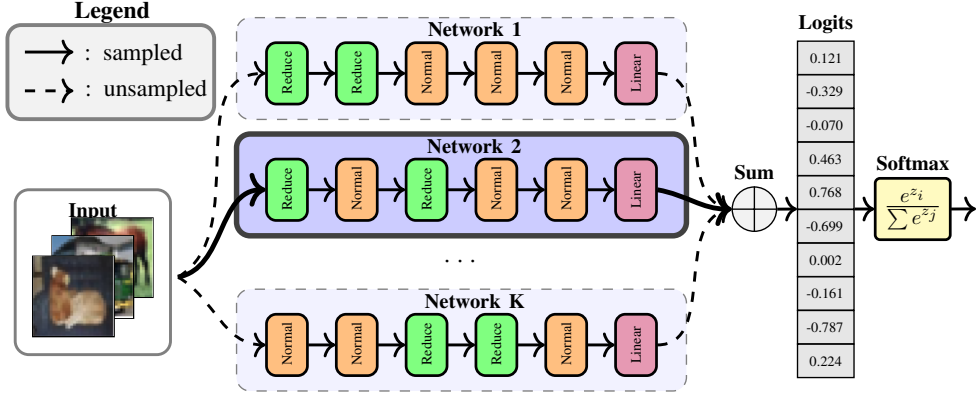}}
    \vspace{0.75em}
    \caption{\sl The Gumbel-softmax trick applied to searching cell location. We randomly sample a candidate network from categorical variable $\bm{\beta}$ and only train sampled network for a given batch. In this example, our algorithm samples Network 2 and update parameters in sampled network and categorical variable $\bm{\beta}$.}
    \vspace{1em}
    \label{fig: reduce location}
\end{figure}

Define a categorical random variable with distribution $\hat{\bm{\beta}}$ and encoded by random one-hot vectors $\rvg \in \{0, 1\}^K$. We construct a ``super'' network as shown in Figure~\ref{fig: reduce location}. Let $f_i$, where $i \in \{1,2,\ldots, K\}$ be a function parameterized by weights $\rvw_i$; we can imagine $f_i$ to represent candidate networks with different locations of \reduce cells. The output $F$ computes the linear combination $F = g_1f_1 + \ldots + g_kf_k$. Intuitively, $F$ samples one branch according to $\rvg$. 

One can imagine learning the optimal indicator vector $\bm{\beta}$ via gradient descent; unfortunately, the sampling operation is not differentiable. Therefore, we leverage the Gumbel-softmax trick~\cite{jang2016categorical}. 
During the forward pass, we sample a one-hot vector according to the formula:
\begin{equation}
    \rvg = \text{one-hot}(\argmax_{i \in \{1,2,\ldots,K\}} {G_i + \log(\hat{\beta}_i)})
    \label{eq: gumble max trick}
\end{equation}
where $G_i \sim \text{Gumbel}(0, 1)$ i.i.d samples drawn from the standard Gumbel distribution. During the backward pass, we use the straight-through Gumbel softmax estimator which replaces $\rvg$ with $\tilde{\rvg}$ during the gradient update:
\begin{equation}
    \tilde{g}_i = \frac{\exp{(\log{g_i} + G_i) / \tau}}{\sum_{k}(\exp{(\log g_k + G_k)})/\tau}
    \label{eq: gumbel softmax approximation}
\end{equation}
where $\tau$ is a temperature parameter. The parameter $\tau$ controls the sharpness of the softmax approximation; $\tilde{\rvg}=\rvg$ as $\tau \to 0$, whereas $\tilde{\rvg}$ becomes an uniform distribution as $\tau \to \infty$. 

Equipped with a fully differentiable technique for learning the parameter $\bm{\beta}$, we now train both the network parameters $\rvw$ and categorical parameter $\bm{\beta}$. Given a train ($T$) and validation ($V$) dataset of labeled pairs $(x, y)$ drawn from a joint distribution $(X,Y)$, Algorithm~\ref{alg: reduce cells loc} update $\rvw$ and $\bm{\beta}$ in alternative fashion to estimate following bilevel optimization problem: 
\begin{equation}
\min_{\bm{\beta} \in \R^K} \:\:\:\:\: \sum_{(x, y) \in V} \mathcal{L}(F_{\rvw^*, \bm{\beta}}(x), y) \:\:\:\: \text{s.t. } \:\:\:\:\: \rvw^* = \argmin_{\rvw \in \R^d} \sum_{(x, y) \in T} \mathcal{L}(F_{\rvw, \bm{\beta}}(x), y)
\end{equation}
Once the training terminates, we select the positions of the \reduce cells in the final network looking at the peak achieved by the categorical distribution (without any Gumbel sampling):
\begin{equation}
    \rvg = \text{one-hot}(\argmax_{i \in \{1,2,\ldots K\}} {\log(\hat{{\beta}}_i)})
    \label{eq: pick candidate}
\end{equation}

\section{Evaluations}
\label{sec: Evaluations}

We now present a series of evaluation studies showing (i) superior performance of \algname{} with state-of-the-art methods for private inference, (ii) transferability of \algname{} cells to more complex datasets (Tiny-ImageNet and ImageNet), (iii) thorough ablation studies that describe the impact of each component in \algname{}.

\subsection{Experimental setup}
\label{subsec: experimental setup}

\algname{} adopts the \Delphi{} PI protocol described in Section~\ref{sec: pibackground}. 
Our reported online runtime includes total computation and communication costs from both the client and server. 
We evaluate \algname{} on CIFAR-100~\cite{CIFAR100}, Tiny-ImageNet~\cite{Le2015TinyIV}, and ImageNet~\cite{ILSVRC15}. Datasets are preprocessed only with image normalization, random horizontal flips, padding, and random crops for the fair comparison. We use NVIDIA Quadro RTX8000 for network training and an Intel i9-10900X CPU running at 3.70GHz with 64GB of memory for benchmarking PI runtime.
We include details on the exact hyperparameter setup for searching cell structure and location in Appendix~\ref{appendix: hyperparameters protocols}.

\subsection{Experimental results and ablation studies}

\noindent\textbf{The \algname{} search space is very efficient in terms of ReLUs.} We compare the network performance between our \algname{} space with (a variant of) the regular DARTS search space given similar ReLU budgets. 
To save ReLUs for both methods, we use a technique called \emph{ReLU sharing} inspired from the ReLU shuffling method in \cite{NEURIPS2020_c519d47c}. If a node connects to two or more convolution modules, convolution modules share a ReLU pre-computed feature map from the input. This reduce the ReLU operations without changing the functionality. We defer a visual explanation of ReLU sharing in Figure~\ref{fig: relu sharing} from Appendix~\ref{appendix: sphynx vs. micro search space}. 
With all this, the micro-search space still faces the limitation of scaling down to a deficient ReLU budget network. The smallest network we could design using the micro-search space involved 78K ReLU counts with channel count $C{=}1$ and depth $D{=}4$. Furthermore, \algname{} achieves 69.57\% test accuracy with 50K ReLUs ($C{=}5, D{=}10$) while the network discovered using the micro-search space only achieves 47.24\% test accuracy with 78K ReLUs. We include more detailed comparisons for various ReLU budgets in Appendix~\ref{appendix: sphynx vs. micro search space}
\begin{figure}[!ht]
\centering
    \resizebox{0.45\textwidth}{!}{
    \begin{tikzpicture}[]
  
\begin{axis}  
[  
    ybar,  
    enlargelimits=0.15,  
    height=6.5cm,
    ylabel={C100 Test Acc.}, 
    ylabel style={at={(-0.07, 0.5)}},
    xlabel={ReLU Budget},  
    symbolic x coords={25K, 30K, 40K, 50K}, 
    xtick=data,  
    nodes near coords, 
    nodes near coords align={vertical},  
	legend style={at={(0.5,1.11), scale=0.5},
	anchor=north,legend columns=-1},
    ]  
\addplot+[error bars/.cd, y fixed, y dir=both, y explicit]
        coordinates {   (25K, 66.13) +- (0, 0.09)
                        (30K, 67.37) +- (0, 0.05)
                        (40K, 68.23) +- (0, 0.21)
                        (50K, 69.57) +- (0, 0.06)
                        };  

\addplot+[error bars/.cd, y fixed, y dir=both, y explicit] 
        coordinates {   (25K, 61.88) +- (0, 0.22)
                        (30K, 62.38) +- (0, 0.27)
                        (40K, 65.96) +- (0, 0.12)
                        (50K, 67.02) +- (0, 0.91)
                        };  


\legend{Reduce Loc. Search., (1/3, 2/3) Reduce Loc.}
\end{axis}  
\end{tikzpicture}  }
    \vspace{1em}
    \caption{\sl Networks with Algorithm~\ref{alg: reduce cells loc} enjoys the performance boost over the conventional $D/3$ and $2D/3$ \textit{reduce} cells approaches over all range of ReLU budget. Each plot includes mean and standard deviation from three different random seed.} \vspace{-0.2em}
    \label{fig: reduce loc search vs default}

\end{figure}
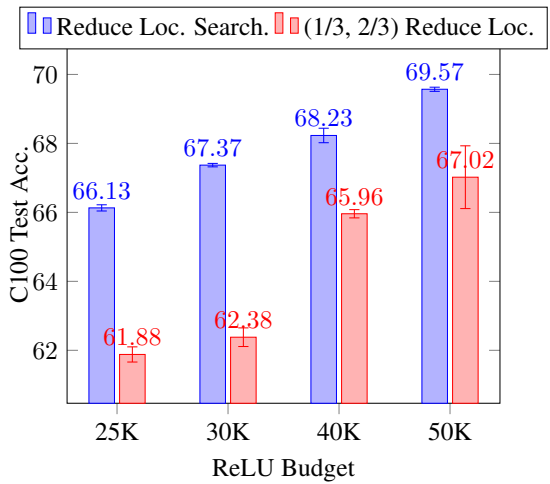

\noindent\textbf{Optimizing locations of \textit{reduce} cells is necessary.} 
We first compare between \textit{reduce} cell location optimized network and $D/3$ and $2D/3$ fixed networks. 
We collect four different ReLU budget networks with three different random seeds: 25K, 30K, 40K, and 50K ReLU. We observe the network performance, especially at lower ReLU budgets, enjoys the benefit of Algorithm~\ref{alg: reduce cells loc} with huge gains in test accuracy. The 25K ReLU network achieves $66.13\%$ test accuracy from Algorithm~\ref{alg: reduce cells loc} while fixing locations to $D/3$ and $2D/3$ only gives $61.88\%$. As shown in Figure~\ref{fig: reduce loc search vs default}, the networks uniformly enjoy boosts in accuracy via Algorithm~\ref{alg: reduce cells loc} in various ReLU budget networks.

The natural follow-up question is whether Algorithm~\ref{alg: reduce cells loc} is capable of finding optimal locations of \textit{reduce} cells. We conduct experiments on CIFAR-100 with three different ReLU budgets: 25K, 30K, and 40K, where initial channels fixed at $C{=}5$ and depths chosen as $D{=}5,6,\text{and}\:8$, respectively. We directly compare the output $\hat{\bm{\beta}}$ from Algorithm~\ref{alg: reduce cells loc} with test accuracy from all possible candidates through a grid search. Our empirical results show that our approach capable of finding the best \textit{reduce} cell locations, which maximizes the test accuracy. We note that our methods can save computational resources over the grid search by a factor of $\frac{\text{\# of candidates}}{2}$ under the assumption that Algorithm~\ref{alg: reduce cells loc} and final network training spends equivalent computational budgets (location searching + final network training). We note that we use the same number of epochs in Algorithm~\ref{alg: reduce cells loc} for the final network training. 

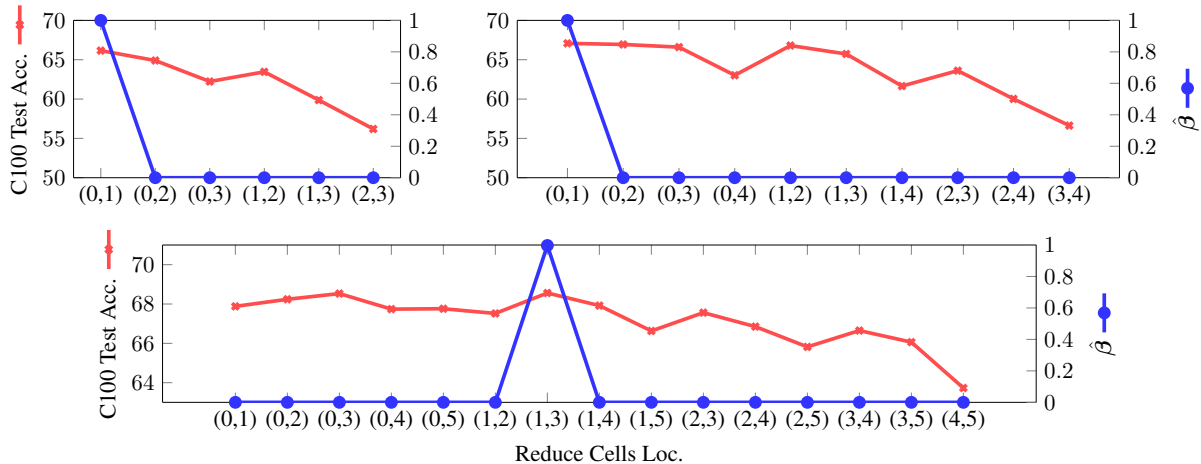
\begin{figure}[!t]
    \centering
    \resizebox{\textwidth}{!}{
    \begin{tabular}{c c}
         \begin{tikzpicture}
\pgfplotsset{
}

\begin{axis}[
  width=0.4\textwidth,
  height=4cm,
  axis y line*=left,
  ymin=50, ymax=70,
  ylabel={C100 Test Acc. \ref{plot_one}},
  symbolic x coords = {{(0,1)}, {(0,2)}, {(0,3)}, {(1,2)}, {(1,3)}, {(2,3)}},
  xtick=data
]
\addplot[mark=x,red!70, ultra thick]
  coordinates{
    ({(0,1)}, 66.16)
    ({(0,2)}, 64.90)
    ({(0,3)}, 62.23)
    ({(1,2)}, 63.46)
    ({(1,3)}, 59.87)
    ({(2,3)}, 56.20)
}; \label{plot_one}

\end{axis}

\begin{axis}[
  width=0.4\textwidth,
  height=4cm,
  axis y line*=right,
  axis x line=none,
  ymin=0, ymax=1,
  symbolic x coords = {{(0,1)}, {(0,2)}, {(0,3)}, {(1,2)}, {(1,3)}, {(2,3)}},
]

\addplot[mark=*,blue!80, ultra thick]
  coordinates{
    ({(0,1)}, 0.99997)
    ({(0,2)}, 0.00001)
    ({(0,3)}, 0.00001)
    ({(1,2)}, 0.00001)
    ({(1,3)}, 0.00001)
    ({(2,3)}, 0.00001)
}; \label{plot_two}
\end{axis}

\end{tikzpicture} & \begin{tikzpicture}
\pgfplotsset{
}

\begin{axis}[
  width=0.656\textwidth,
  height=4cm,
  axis y line*=left,
  ymin=50, ymax=70,
  symbolic x coords = {{(0,1)}, {(0,2)}, {(0,3)}, {(0,4)}, {(1,2)}, {(1,3)}, {(1,4)}, {(2,3)}, {(2,4)}, {(3,4)}},
  xtick=data
]
\addplot[mark=x,red!70, ultra thick]
  coordinates{
    ({(0,1)}, 67.08)
    ({(0,2)}, 66.94)
    ({(0,3)}, 66.60)
    ({(0,4)}, 63.02)
    ({(1,2)}, 66.80)
    ({(1,3)}, 65.73)
    ({(1,4)}, 61.65)
    ({(2,3)}, 63.61)
    ({(2,4)}, 60.02)
    ({(3,4)}, 56.63)
}; \label{plot_one2}

\end{axis}

\begin{axis}[
  width=0.656\textwidth,
  height=4cm,
  axis y line*=right,
  axis x line=none,
  ymin=0, ymax=1,
  symbolic x coords = {{(0,1)}, {(0,2)}, {(0,3)}, {(0,4)}, {(1,2)}, {(1,3)}, {(1,4)}, {(2,3)}, {(2,4)}, {(3,4)}},
  ylabel={$\hat{\bm{\beta}}$ \ref{plot_two2}}
]

\addplot[mark=*,blue!80, ultra thick]
  coordinates{
    ({(0,1)}, 0.99997)
    ({(0,2)}, 0)
    ({(0,3)}, 0)
    ({(0,4)}, 0)
    ({(1,2)}, 0)
    ({(1,3)}, 0)
    ({(1,4)}, 0)
    ({(2,3)}, 0)
    ({(2,4)}, 0)
    ({(3,4)}, 0)
}; \label{plot_two2}
\end{axis}

\end{tikzpicture} \\
         \multicolumn{2}{c}{\begin{tikzpicture}
\pgfplotsset{
}

\begin{axis}[
  width=15cm,
  height=4cm,
  axis y line*=left,
  ymin=63, ymax=71,
  xlabel=Reduce Cells Loc.,
  ylabel={C100 Test Acc. \ref{plot_one}},
  symbolic x coords = {{(0,1)}, {(0,2)}, {(0,3)}, {(0,4)}, {(0,5)}, {(1,2)}, {(1,3)}, {(1,4)}, {(1,5)}, {(2,3)}, {(2,4)}, {(2,5)}, {(3,4)}, {(3,5)}, {(4,5)}},
  xtick=data
]
\addplot[mark=x,red!70, ultra thick]
  coordinates{
    ({(0,1)}, 67.88)
    ({(0,2)}, 68.24)
    ({(0,3)}, 68.53)
    ({(0,4)}, 67.74)
    ({(0,5)}, 67.76)
    ({(1,2)}, 67.52)
    ({(1,3)}, 68.56)
    ({(1,4)}, 67.92)
    ({(1,5)}, 66.63)
    ({(2,3)}, 67.56)
    ({(2,4)}, 66.85)
    ({(2,5)}, 65.82)
    ({(3,4)}, 66.65)
    ({(3,5)}, 66.06)
    ({(4,5)}, 63.73)
}; \label{plot_one3}

\end{axis}

\begin{axis}[
  width=15cm,
  height=4cm,
  axis y line*=right,
  axis x line=none,
  ymin=0, ymax=1,
  symbolic x coords = {{(0,1)}, {(0,2)}, {(0,3)}, {(0,4)}, {(0,5)}, {(1,2)}, {(1,3)}, {(1,4)}, {(1,5)}, {(2,3)}, {(2,4)}, {(2,5)}, {(3,4)}, {(3,5)}, {(4,5)}},
  ylabel={$\hat{\bm{\beta}}$ \ref{plot_two3}}
]

\addplot[mark=*,blue!80, ultra thick]
  coordinates{
    ({(0,1)}, 0)
    ({(0,2)}, 0)
    ({(0,3)}, 0)
    ({(0,4)}, 0)
    ({(0,5)}, 0)
    ({(1,2)}, 0)
    ({(1,3)}, 0.997)
    ({(1,4)}, 0)
    ({(1,5)}, 0)
    ({(2,3)}, 0)
    ({(2,4)}, 0)
    ({(2,5)}, 0)
    ({(3,4)}, 0)
    ({(3,5)}, 0)
    ({(4,5)}, 0)
}; \label{plot_two3}
\end{axis}

\end{tikzpicture}}
    \end{tabular}}
    \vspace{0.5em}
    \caption{Comparison on learned $\hat{\bm{\beta}}$ and a grid search on all possible locations of reduce cells' test accuracy. The x-axis represents the locations of two reduce cells. Left and right y-axis represents the CIFAR-100 test accuracy and $ \hat{\bm{\beta}}$ respectively. Categorical parameter $\bm{\beta}$ matches to the best performing network candidate from the grid search. \textit{Upper-Left}: 25K ReLU budget network with depth 5. \textit{Upper-Right}: 30K ReLU budget network with depth 6. \textit{Bottom}: 40K ReLU budget network with depth 8.}
    \label{fig: reduce cell alg vs grid search}
\end{figure}

\begin{figure}[!t]
    \centering
    \resizebox{0.9\textwidth}{!}{
    \begin{tabular}{c c}
         \begin{tikzpicture}[trim left={(-0.85, -0.5)}, trim right={(7.0, 0.5)}]
    \begin{semilogxaxis}[legend pos=north west,
        legend style={nodes={scale=0.62, transform shape}},
        width=0.6\textwidth,
        height=5cm,
        xlabel= Online Runtime (s),
        ylabel=Test Accuracy (\%),
        xlabel style={at={(0.5, -0.07)}},
        ylabel style={at={(-0.05, 0.5)}},
        title=Pareto Curve on CIFAR-100,
        title style={at={(0.5, 0.92)}},
        xmin=0, xmax=10,
        ymin=62,ymax=77,
        axis background/.style={fill=blue!3},
        grid=both,
        log basis x = 2,
        /pgf/number format/1000 sep={\,},
        log ticks with fixed point,
        grid style={line width=.1pt, draw=gray!10},
        major grid style={line width=.2pt,draw=gray!50},
        ]
    \addplot[mark=*,blue, ultra thick] 
        plot coordinates {
            (0.727, 66.23)
            (0.861, 67.43) 
            (1.105, 68.39)
            (1.335, 69.64)
            (1.805, 71.17)
            (2.385, 72.97)
            (5.120, 75.15)
        };
    \addlegendentry{\algname{}}
    

    \addplot[mark=x,red, ultra thick] 
        plot coordinates {
            (1.670, 63.6)
            (2.000 , 68.1)
            (2.300, 68.7)
            (7.500, 75.5)
        };
    \addlegendentry{CryptoNAS}

    \addplot[mark=square,green, ultra thick] 
        plot coordinates {
            (.579, 66.0)
            (.738, 64.36)
            (1.190, 67.8)
            (2.180, 71.1)
        };
    \addlegendentry{DeepReDrop}

    \addplot[mark=diamond,orange, ultra thick] 
        plot coordinates {
            (1.230, 66)
            (4.440, 67)
            (6.500, 68)
        };
    \addlegendentry{DELPHI}
    
    \addplot[mark=triangle, gray, ultra thick] 
        plot coordinates {
            (2.22, 67.5)
        };
    \addlegendentry{SAFENet}
    \end{semilogxaxis}
\end{tikzpicture} & \begin{tikzpicture}[trim left={(-0.85, -0.5)}, trim right={(7.0, 0.5)}]
    \begin{semilogxaxis}[legend pos=south east,
        legend style={nodes={scale=0.8, transform shape}},
        width=0.6\textwidth,
        height=5cm,
        xlabel= Online Runtime (s),
        ylabel=Test Accuracy (\%),
        xlabel style={at={(0.5, -0.07)}},
        ylabel style={at={(-0.05, 0.5)}},
        title=Pareto Curve on ImageNet,
        title style={at={(0.5, 0.9)}},
        xmin=0, xmax=200,
        ymin=30,ymax=80,
        axis background/.style={fill=blue!3},
        grid=both,
        log basis x = 2,
        /pgf/number format/1000 sep={\,},
        log ticks with fixed point,
        grid style={line width=.1pt, draw=gray!10},
        major grid style={line width=.2pt,draw=gray!50},
        ]
    \addplot[mark=*,blue, ultra thick] 
        plot coordinates {
            (3.67, 38.23)
            (7.34, 51.85)
            (10.98, 57.72)
            (14.67, 60.656)
            (18.34, 64.43)
            (22.02, 66.00)
        };
    \addlegendentry{\algname{}}

    \addplot[mark=x,red, ultra thick] 
        plot coordinates {
            (6.06, 31.93)
            (12.12, 47.09)
            (24.23, 60.15)
            (48.45, 69.76)
        };
    \addlegendentry{ResNet18}

    \addplot[mark=square,green, ultra thick] 
        plot coordinates {
            (9.87, 39.68)
            (19.74, 49.76)
            (155.37, 70.37)
        };
    \addlegendentry{VGG11}

    \end{semilogxaxis}
\end{tikzpicture} \\
    \end{tabular}}
    \vspace{0.5em}
    \caption{\textit{Left:} Pareto curve on \algname{} compare to state-of-the-art PI methods on CIFAR-100. \textit{Right:} Pareto curve on \algname{} compare to VGG11 and ResNet18 via \Delphi{} protocol. Online runtimes include both client and server online cost.}
    \label{fig: C100 pareto curve}
\end{figure}
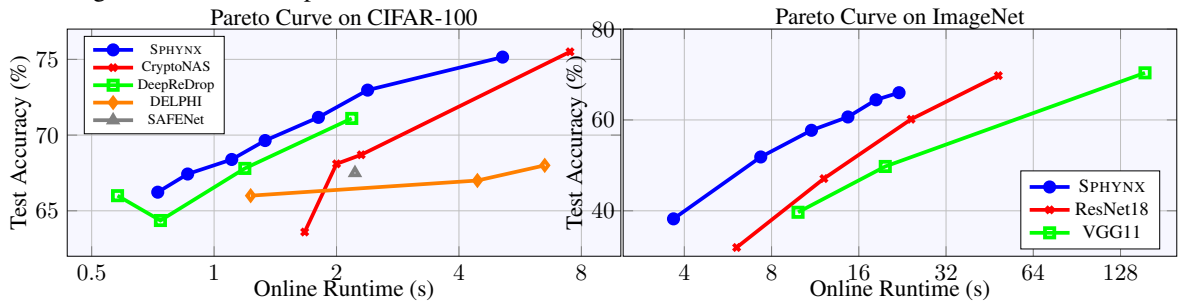

\noindent\textbf{\algname{} achieves the accuracy-latency Pareto frontier for CIFAR-100.} Using \algname{}, we design a range of ReLU-budgeted networks to understand the impact of accuracy-latency tradeoffs in PI. We plot Pareto curves in Figure~\ref{fig: C100 pareto curve} (\algname{} in blue) comparing with prior PI methods: CryptoNAS~\cite{NEURIPS2020_c519d47c}, \Delphi{}~\cite{DELPHI}, DeepReDuce~\cite{DEEPREDUCE}, and SAFENet~\cite{SAFENET}. \algname{} strictly outperforms CryptoNAS, \Delphi, and SAFENet for the entire range of latency models. \algname{} achieves 66.13\% test accuracy with 2.3\texttimes{} faster online latency than CryptoNAS models with 63.60\%, and 1.7\texttimes{} faster than \Delphi{} with 66.00\%. \textcolor{black}{We approximate the online latency from SAFENet (67.50\% test acc.) due to different specs in hardware. We leverage the our latency (4.4s) from \Delphi{} model with 67.00\% and the Delphi latency (14.4s) with 67.3\% model from SAFENet and estimate SAFENet online latency to be 2.2s with our hardware setup}. Moreover, \algname{} achieves the par test accuracy to SAFENet with 2.6\texttimes{} faster PI latency.

\algname{} achieves better results over DeepReDuce in all ranges of online runtime, except at very low-ReLU budgets (24.6K). Our smallest network (25.6K ReLU) is on par with DeepReduce in test accuracy but 25\%\texttimes{} slower in online runtime. 
This is probably due to the fact that \algname{} contains more linear layers than 24.6K DeepReDuce network resulting a viable difference on PI latency, especially in an extreme low-ReLU regime. Apart from this, \algname{} dominates as a Pareto frontier on all range of DeepReDuce case\footnote{We note that our comparison here includes results without knowledge distillation (KD). The original DeepReduce paper empirically showed performance boosts in test accuracy using KD with a high-ReLU network as teacher. For brevity, we defer additional \algname{} results with KD in Appendix~\ref{appendix: hyperparameters protocols} and have observed similar trends to the Pareto curve in Figure~\ref{fig: C100 pareto curve} between \algname{} and DeepReDuce.}.

 \begin{table}[!t]
 \centering
    \resizebox{0.62\textwidth}{!}{
    \begin{tabular}{|c | c | c | c | c | c|}
    \hline
    \textbf{\algname{}} & \multicolumn{5}{c|}{\textbf{Tiny-ImageNet} (ResNet18 (2228K), Acc=61.28\%)} \\
    \hline
    {ReLU} \# & 102.4K & 204.8K & 286.7K & 491.5K & 614.4K \\
    \hline
    {PI Lat.} & 2.35s & 4.40s & 6.14s & 10.2s & 12.5s \\
    {Test Acc.} & 48.44\% & 53.51\% & 56.72\% & 59.18\% & 60.76\% \\
    \hline
    \end{tabular}}
    \vspace{0.8em}
    \caption{\sl \algname{} Tiny-ImageNet results via transfer learning from cells found from CIFAR-100. \algname{} with 614K ReLU achieves roughly par test accuracy to ResNet18 with 2228K ReLU counts.}
    \vspace{0em}
\end{table}


\noindent\textbf{Transfer learning from CIFAR-100 cells to Tiny-ImageNet and ImageNet} Next, we show that the \algname{} cells learned from CIFAR-100 transfer to more complicated datasets such as Tiny-ImageNet and ImageNet. In case of Tiny-ImageNet, we found that an apples-to-apples comparison with DeepReDuce is difficult since only KD results were reported. In particular, the teacher model in DeepReDuce (ResNet18 with 2228K ReLUs) achieves 61.28\% test accuracy while the student model surpasses it (e.g., 917K ReLU student network achieves 64.66\% w/ KD). Under the assumption that the student models without KD are less potent than the teacher model, \algname{} with 614K ReLUs achieves competitive test accuracy 60.57\%, roughly par with ResNet18 with 2224K ReLUs, with over 0.25\texttimes{} fewer ReLUs.


We also transfer \algname{} CIFAR-100 cells to ImageNet to verify the promise of \algname{} for large-scale complex datasets. 
We adopt a slightly different network architecture for ImageNet following DARTS; details are in Appendix~\ref{appendix: architecture details for imagenet}. 
\textcolor{black}{Since the \Delphi{} protocol fails in the large-scale dataset, we estimate the PI latency by summing average GC latency per ReLU from our Tiny-ImageNet models, and plaintext inference time as an approximation to linear layer computation}. \algname{} strictly dominates VGG11 and ResNet18 for the entire range of latency values. 

\noindent\textbf{Additional ablation studies.} We further support our approach by providing several other ablation studies. Due to space constraints, we only summarize these results and defer details to Appendix~\ref{appendix: additional ablation studies}. (i) The network performance deviates minuscule from various selections in initial channels and depths given a ReLU budget. (ii) ReLU balancing outperforms FLOP balancing. (iii) Cells with $N{=}7$ (more linear operations) found from \algname{} achieve superior performance over cells with $N{=}5$ (less linear operations). (iv) 
Using cells reported from DARTS/PC-DARTS but modify as described in Section~\ref{subsec: Efficient search space} 
is significantly worse compare to using cells found from the \algname{} framework.

\section{Related Work and Discussion}
\label{sec: related works}


CryptoNets~\cite{gilad2016cryptonets} is one of the earliest works on PI, fully leveraging homomorphic encryption (FHE) to guarantee data (but not model) privacy. However, CryptoNets only allows polynomial activations due to reliance on fully homomorphic encryption. Subsequent works, including MiniONN~\cite{liu2017oblivious}, SecureML~\cite{mohassel2017secureml}, Gazelle~\cite{juvekar2018gazelle}, and \Delphi{}~\cite{DELPHI}, have focused on providing both data and model privacy and support standard nonlinear activation functions such as ReLUs. These approaches isolate linear and non-linear operations and apply different protocols to each. Several state-of-the-art PI protocols leverage expensive Garbled Circuits for ReLU operations. Consequently, subsequent works have concentrated on reducing ReLU operations, such as ReLU approximation~\cite{DELPHI, SAFENET}, ReLU-efficient network design via NAS~\cite{NEURIPS2020_c519d47c}, and pruning ReLU layers~\cite{DEEPREDUCE}. A separate line of PI literature, including DeepSecure~\cite{rouhani2018deepsecure} and XONN~\cite{riazi2019xonn}, leverages binarized neural networks but these approaches underperform in test accuracy than conventional networks. 

Early NAS approaches leveraged RL-based controllers~\cite{zoph2016neural} to design entire convolutional networks. Macro-search methods adopt a simple chain-structure skeleton with skip-connections between layers since the search complexity tends to exponentially increase depending on the number of design components. NASNet~\cite{zoph2018learning} proposes transforming macro-search tasks into finding block modules, which can be repeatedly stacked up to form the final architecture. NASNet further improves the performance over \cite{zoph2016neural}; however, both approaches consumed substantial computational resources, running into thousands of GPU-days. Subsequent NAS works on micro-search space have focused on reducing the search time with techniques such as gradient-based optimization~\cite{cai2018proxylessnas, liu2018darts, xie2018snas, Xu2020PC-DARTS:, dong2019search}, weight-sharing super-network~\cite{pham2018efficient, bender2018understanding, li2019random}, evolutionary algorithms~\cite{real2018regularized, liu2018hierarchical}, and hyperparameter optimization techniques~\cite{falkner2018bohb, li2019random}. We envision that \textsc{Sphynx} extends to any of these methods.


Numerous follow-up avenues still persist.  \algname{} finds the network in three stages that are somewhat decoupled: (i) find cells; (ii) choose the initial channels and depth given a ReLU budget, and (iii) optimize for the location of the reduce cells. Integrating these three phases into a single technique might improve final performance by reducing the gap between the search and evaluation phase~\cite{Yang2020NAS}. Moreover, the objective function in \algname{} is the standard empirical risk used in training. The impact of incorporating additional terms (such as ReLU counts) to the objective function will be an interesting direction for future work.




{{
\bibliographystyle{unsrt}
\bibliography{neurips_biblio}
}}




\appendix
\section{Background on DARTS: Differentiable Architecture Search}
\label{appendix: darts background}

This section describes the micro-search NAS optimization problem, and briefly discusses DARTS details.

In the micro-search space, the neural architecture has a skeleton of repeatedly stacked structures or modules denoted as cells. Following the pioneer works from NASNet~\cite{zoph2018learning}, searching the architecture candidate is commensurate to search two types of cells: normal cells and reduce cells. 
Normal cells learn the the high-level features returning the exact spatial resolution to the input. Conversely, reduce cells extract the high-level representations and reduce the spatial resolution (commonly reduced by half on height and width). Since the final architecture is defined as stacked normal and reduce cells, constructing the architecture is equivalent to selecting normal/reduce cells in the micro-search space based NAS perspective. 

In detail, consider a supervised learning NAS setup. We have a train ($T$) and validation ($V$) dataset of labeled pairs $(x, y)$ drawn from a joint distribution $(X, Y)$. Let $f_{\rvw, a}$ be the candidate network parameterized by $\rvw$ given a network constructed with normal/reduce cells from discrete architecture space $a \in A$. Finding the normal/reduce cells $a \in A$ is equivalent to solving following combinatorial optimization problem:
\begin{equation}
    \min_{a \in A} \sum_{(x, y) \in V} \mathcal{L}(f_{\rvw^*, a}(x), y) \:\:\:\: \text{s.t.} \:\:\:\:\: \rvw^* = \argmin_{\rvw \in \R^d} \sum_{(x, y) \in T} \mathcal{L}(f_{\rvw, a}(x), y) \\
\end{equation}

Gradient-based NAS approaches including DARTS require continuous relaxation step of discrete architecture space. Let $\Theta$ be some continuous relaxation of the discrete architecture space $A$. Then the gradient-based approaches solve the following objectives:
\begin{equation}
    \min_{\theta \in \Theta} \sum_{(x, y) \in V} \mathcal{L}(f_{\rvw^*, \theta}(x), y) \:\:\:\: \text{s.t.} \:\:\:\:\: \rvw^* = \argmin_{\rvw \in \R^d} \sum_{(x, y) \in T} \mathcal{L}(f_{\rvw, \theta}(x), y) \\
    \label{eq: continuous relaxation on search}
\end{equation}
and derive a final architecture by mapping back to discrete space: \texttt{Sample:}$\Theta \to A$ (e.g., magnitude based selection).

During the search phase, DARTS involves a continuous relaxation of the discrete cell space via weighted summation of all possible operation between node $i$ and node $j$ by leveraging softmax. From an operation set $\mathcal{O}$, the continuous relaxed operation $\bar{o}$ is expressed with learnable architecture parameter $\theta$:
\begin{equation}
    \bar{o}^{(i,j)}(z^{(i)}) = \sum_{o \in \mathcal{O}} \frac{\exp{(\theta_o^{(i,j)})}}{\sum_{o^\prime \in \mathcal{O}} \exp{(\theta_{o^\prime}^{(i,j)})}} o(z^{(i)})
\end{equation}
where $\bm{\theta} = \{\theta^{(i, j)}\}$ and $\theta^{(i, j)} \in \R^{|\mathcal{O}|}$. During the search phase, DARTS alternatively update the network parameter weights $\rvw$ and architecture parameter $\bm{\theta}$ via gradient descent algorithms. 
Finally, DARTS selects the final normal/reduce cells with a function \texttt{Sample} selecting two strongest predecessors for each intermediate node from architecture parameter $\bm{\theta}$ by $\max_{o \in \mathcal{O}} \frac{\exp({\theta_o^{(i,j)}})}{\sum_{o^\prime \in \mathcal{O}, o \neq zero}\exp{(\theta_{o^\prime}^{(i, j)})}}$. 
\section{Found cells from \algname{}}
\label{appendix: cell pictures}

\begin{figure}[!ht]
    \centering
    \renewcommand{\arraystretch}{0.2}
    \def\sw{0.45\linewidth}
    \begin{tabular}{c c}
        \includegraphics[width=\sw]{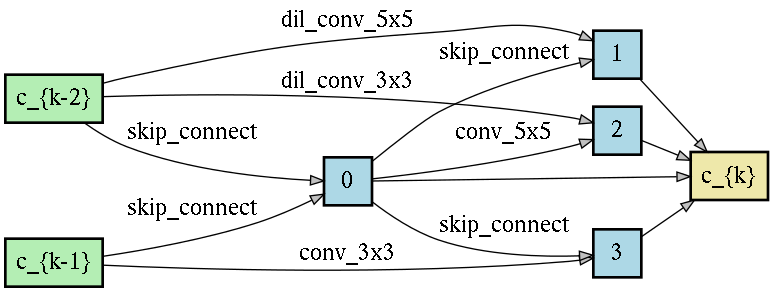} &
        \includegraphics[width=\sw]{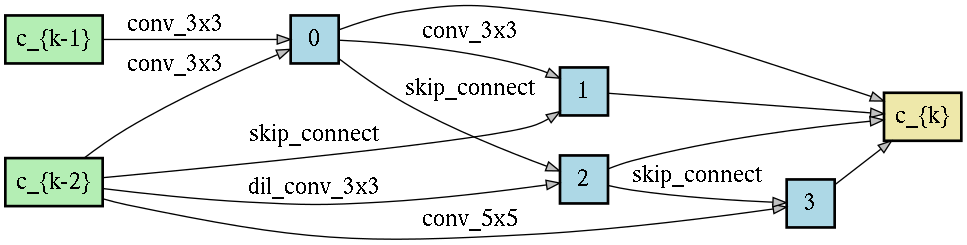} \\
        \algname{}-Normal Cell & \algname{}-Reduce Cell \\
    \end{tabular}
    \vspace{1.5em}
    \caption{\textit{Normal} and \textit{reduce} cells found by \algname{} from CIFAR-100.}
    \vspace{1.5em}
    \label{fig: cells}
\end{figure}

\newpage 
\section{CIFAR-100, Tiny-ImageNet, and ImageNet results in Table}
\label{appendix: cifar100, tiny, imagenet results in table}

\begin{table}[ht]
    \centering 
    \caption{Comparison of \algname{} and existing state-of-the-art methods in private inference on CIFAR-100. Higher the test accuracy and lower the PI latency the better. Our empirical results outperforms existing PI methods in various ReLU budget networks on test accuracy and private inference latency. We run \algname{} with three different random seeds and report mean and standard deviation.}
    \resizebox{\textwidth}{!}{
    \begin{threeparttable}
        \begin{tabular}{c | c c c c | c | c c c c}
            \toprule
             & \multicolumn{1}{c}{\textbf{Methods}} & \multicolumn{1}{c}{\textbf{ReLUs}} & \multicolumn{1}{c}{\textbf{Test Acc.}} & \multicolumn{1}{c|}{\textbf{PI Lat.}} & & \multicolumn{1}{c}{\textbf{Methods}} & \multicolumn{1}{c}{\textbf{ReLUs}} & \multicolumn{1}{c}{\textbf{Test Acc.}} & \multicolumn{1}{c}{\textbf{PI Lat.}}\\
            \midrule
            \multirow{7}{*}{\rotatebox[origin=c]{90}{$\text{ReLU} \leq 55K$}} & \multicolumn{1}{c}{\algname{}} & \multicolumn{1}{c}{25.6K} & $\bm{66.13{\pm}0.09}$\% & 727ms & \multirow{7}{*}{\rotatebox[origin=c]{90}{$\text{ReLU} \leq 350K$}} & \multicolumn{1}{c}{\algname{}} & \multicolumn{1}{c}{71.7K} & $\bm{71.06{\pm}0.15}$\% & \textbf{1805}ms \\
             & \multicolumn{1}{c}{DeepReDuce} & \multicolumn{1}{c}{24.6K} & 66.00\% & \textbf{579}ms &  & \multicolumn{1}{c}{CryptoNAS} & \multicolumn{1}{c}{86.0K} & 68.13\% & 2000ms \\
             & \multicolumn{1}{c}{\algname{}} & \multicolumn{1}{c}{30.2K} & $67.37{\pm}0.05$\% & 861ms &  & \multicolumn{1}{c}{\algname{}} & \multicolumn{1}{c}{102.4K} & $\bm{72.90{\pm}0.06}$\% & \textbf{2385}ms \\
             & \multicolumn{1}{c}{\algname{}} & \multicolumn{1}{c}{41.0K} & $\bm{68.23{\pm}0.21}$\% & \textbf{1105}ms &  & \multicolumn{1}{c}{CryptoNAS} & \multicolumn{1}{c}{100.0K} & 68.30\% & 2300ms \\
             & \multicolumn{1}{c}{\algname{}} & \multicolumn{1}{c}{51.2K} & $\bm{69.57{\pm}0.06}$\% & 1335ms &  & \multicolumn{1}{c}{\Delphi{}} & \multicolumn{1}{c}{180.0K} & 67.00\% & 4440ms \\
             \cline{7-10} & \multicolumn{1}{c}{DeepReDuce} & \multicolumn{1}{c}{49.2K} & 67.80\% & 1190ms &  & \multicolumn{1}{c}{\algname{}} & \multicolumn{1}{c}{230.0K} & $74.93\pm0.16$\% & \textbf{5120}ms \\
             & \multicolumn{1}{c}{\Delphi{}} & \multicolumn{1}{c}{50.0K} & 66.00\% & 1230ms &  & \multicolumn{1}{c}{\Delphi{}} & \multicolumn{1}{c}{300.0K} & 68.00\% & 6500ms \\
             & \multicolumn{1}{c}{CryptoNAS} & \multicolumn{1}{c}{50.0K} & 63.60\% & 1670ms &  & \multicolumn{1}{c}{CryptoNAS} & \multicolumn{1}{c}{344.0K} & 75.64\% & 7500ms \\
             
            \midrule
        \end{tabular}
    \end{threeparttable}}
    \label{table: cifar100}
\end{table}

\begin{table}[ht]
    \centering 
    \caption{\textbf{Tiny-ImageNet and ImageNet results}. No prior art on PI experiment on Tiny-ImageNet and ImageNet except DeepReDuce~\cite{DEEPREDUCE} only on Tiny-ImageNet. We note that apple-to-apple comparison between \algname{} and DeepReDuce is challenging due to KD incorporated in DeepReDuce training. \algname{} strictly outperforms scaled down VGG11 and ResNet18 with less ReLU budget. Our ImageNet training methods on \algname{}, VGG11, and ResNet18 are equivalent as described in Appendix~\ref{appendix: hyperparameters protocols}.}
    \resizebox{0.95\textwidth}{!}{
    \begin{threeparttable}
        \begin{tabular}{c | c c c c| c | c c c c}
            \toprule
             & \multicolumn{1}{c}{\textbf{Methods}} & \multicolumn{1}{c}{\textbf{ReLUs}} & \multicolumn{1}{c}{\textbf{Test Acc.}} & \multicolumn{1}{c|}{\textbf{PI Lat.}}& & \multicolumn{1}{c}{\textbf{Methods}} & \multicolumn{1}{c}{\textbf{ReLUs}} & \multicolumn{1}{c}{\textbf{Test Acc.}} & \multicolumn{1}{c}{\textbf{PI Lat.}}\\
            \midrule
            \multirow{8}{*}{\rotatebox[origin=c]{90}{\textbf{Tiny-ImageNet}}} & \multicolumn{1}{c}{\algname{}} & \multicolumn{1}{c}{102.4K} & 48.44\% & 2350ms & \multirow{8}{*}{\rotatebox[origin=c]{90}{\textbf{ImageNet}}} & \multicolumn{1}{c}{ResNet18} & \multicolumn{1}{c}{288.4K} & 31.93\% & 6060ms \\
              & \multicolumn{1}{c}{\algname{}} & \multicolumn{1}{c}{204.8K} & 53.51\%  & 4401ms& & \multicolumn{1}{c}{\algname{}} & \multicolumn{1}{c}{345K} & \textbf{51.85}\% & 7340ms \\
              & \multicolumn{1}{c}{\algname{}} & \multicolumn{1}{c}{286.7K} & 56.72\%  & 6140ms & & \multicolumn{1}{c}{VGG11} & \multicolumn{1}{c}{470K} & 39.68\% & 9870ms \\
             & \multicolumn{1}{c}{\algname{}} & \multicolumn{1}{c}{491.5K} & 59.12\%  & 10205ms&  & \multicolumn{1}{c}{\algname{}} & \multicolumn{1}{c}{517K} & \textbf{57.72}\% & 10980ms \\
              & \multicolumn{1}{c}{\algname{}} & \multicolumn{1}{c}{614.4K} & \textbf{60.76}\%  & 12548ms& & \multicolumn{1}{c}{ResNet18} & \multicolumn{1}{c}{577K} & 47.09\% & 12120ms \\
              & \multicolumn{1}{c}{ResNet18} & \multicolumn{1}{c}{2228.2K} & 61.28\% & N/A & & \multicolumn{1}{c}{\algname{}} & \multicolumn{1}{c}{862K} & \textbf{64.43}\% & 18340ms  \\
             & \multicolumn{1}{c}{-} & \multicolumn{1}{c}{-} & - & - & & \multicolumn{1}{c}{VGG11} & \multicolumn{1}{c}{940K} & 49.76\% & 19740ms \\
             & \multicolumn{1}{c}{-} & \multicolumn{1}{c}{-} & -  & - & & \multicolumn{1}{c}{\algname{}} & \multicolumn{1}{c}{1034K} & \textbf{66.00}\% & 22020ms \\
             & \multicolumn{1}{c}{-} & \multicolumn{1}{c}{-} & -  & - & & \multicolumn{1}{c}{ResNet18} & \multicolumn{1}{c}{1154K} & 60.15\% & 24225ms \\
            \midrule
        \end{tabular}
    \end{threeparttable}}
    \label{table: tiny and imagenet}
\end{table}

\newpage 
\section{Search/Evaluation Protocols and Hyperparameter Setups}
\label{appendix: hyperparameters protocols}

We provide the search/evaluation training protocol to reproduce the results. This section includes the following hyperparameters setup: 
\begin{itemize}
    \itemsep-0.10em 
    \item Table~\ref{table: cell search protocol.}: Hyperparameter setup to find the \textit{normal/reduce} cells from \algname{} search space. 
    \item Table~\ref{table: location search protocol}: Hyperparameter setup to find the location of reduce cells.
    \item Table~\ref{table: evaluation protocol}: Hyperparameter setup to train the final network on CIFAR-100 and Tiny-ImageNet.
    \item Table~\ref{table: ImageNet evaluation protocol}: Hyperparameter setup to train the final network on ImageNet.
\end{itemize}

\begin{table}[ht]
    \centering
    \caption{\textit{Normal}/\textit{Reduce} cells searching hyperparameters. $\rvw$ and $\bm{\theta}$ stand for network and architecture parameter, respectively.}
    \begin{tabular}{c | c || c | c}
        \hline
        $\rvw$ optimizer & SGD & initial LR & 0.025 \\
        Nesterov & Yes & ending LR & 0.001 \\
        momentum & 0.9 & LR schedule & cosine \\
        $\rvw$ weight decay & 0.0003 & epoch & 50 \\
        batch size & 64 & initial channel & 5 \\
        cells \# & 8 & cutout & No \\
        ops \# & 6 & nodes \# & 7 \\
        random flip & p=0.5 & random crop & Yes \\
        normalization & Yes & grad clip & 5.0 \\
        $\bm{\theta}$ Optim. & Adam & $\bm{\theta}$ init. LR & 0.0003 \\
        $\bm{\theta}$ weight decay & 0.001 & reduce loc. & [$D/3$, $2D/3$] \\
        \hline
    \end{tabular}
    \label{table: cell search protocol.}
\end{table}

\begin{table}[ht]
    \centering
    \caption{\textit{Reduce} cells location search hyperparameters. $\rvw$ and $\bm{\beta}$ stand for network and categorical parameter, respectively.}
    \begin{tabular}{c | c || c | c}
        \hline
        $\rvw$ optimizer & SGD & initial LR & 0.025 \\
        Nesterov & Yes & ending LR & 0.001 \\
        momentum & 0.9 & LR schedule & cosine \\
        $\rvw$ weight decay & 0.0003 & epoch & 600 \\
        batch size & 64 & initial $\tau$ & 1000 \\
        ending $\tau$ & 0.1 & $\tau$ schedule & Linear \\
        random flip & p=0.5 & random crop & Yes \\
        normalization & Yes & grad clip & 5.0 \\
        $\bm{\beta}$ Optim. & Adam & $\bm{\beta}$ init. LR & 0.0003 \\
        $\bm{\beta}$ weight decay & 0.001 & - & - \\
        \hline
    \end{tabular}
    \label{table: location search protocol}
\end{table}

\begin{table}[ht]
    \centering
    \caption{Final network training hyperparameter protocol on CIFAR-100 and Tiny-ImageNet. $D$ stands for the total number of cells in the final network. }
    \begin{tabular}{c | c || c | c}
        \hline
        optimizer & SGD & initial LR & 0.025 \\
        Nesterov & Yes & ending LR & 0 \\
        momentum & 0.9 & LR schedule & cosine \\
        weight decay & 0.0003 & epoch (C100/Tiny) & 600/250 \\
        batch size & 96 & parallel training & No\\
        random flip & p=0.5 & random crop & Yes \\
        normalization & Yes & cutout & No \\
        drop-path prob & 0.2 & aux loss loc. & $2D/3$ \\
        grad clip & 5.0 & aux weight & 0.4 \\
        \hline
    \end{tabular}
    \label{table: evaluation protocol}
\end{table}

\begin{table}[ht]
    \centering
    \caption{Final Network Training Hyperparameter Protocol on ImageNet. $D$ stands for the total number of cells in the final network. }
    \begin{tabular}{c | c || c | c}
        \hline
        optimizer & SGD & initial LR & 0.1 \\
        Nesterov & Yes & Lr schedule & step \\
        momentum & 0.9 & lr decay & 30, 60, 90\\
        lr decay mult. factor & 0.1 & epoch & 120 \\
        weight decay & 0.0003 & \# of GPUs & 4 \\
        batch size & 768 & parallel training & Yes \\
        random flip & p=0.5 & random crop & Yes \\
        normalization & Yes & cutout & No \\
        drop-path prob & 0.0 & aux loss loc. & $2D/3$ \\
        label smoothing & No & aux weight & 0.4 \\
        grad clip & 5.0 & - & - \\
        \hline
    \end{tabular}
    \label{table: ImageNet evaluation protocol}
\end{table}

\section{Detail descriptions in Ablation studies}
\label{appendix: ablation}

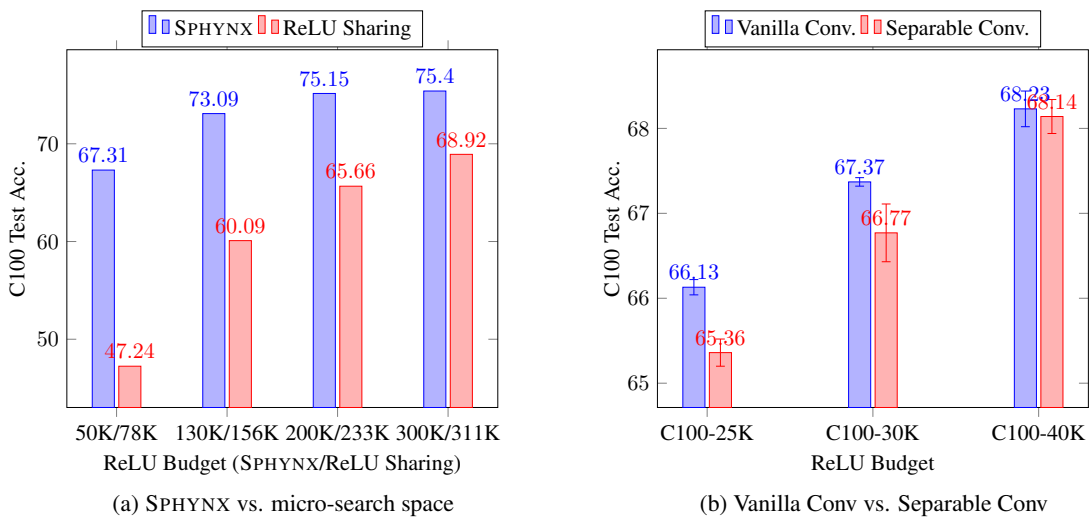
\begin{figure}[!t]
\centering
    \begin{tabular}{c c}
        \resizebox{0.45\textwidth}{!}{\begin{tikzpicture}[trim left={(-1.0,0)}, trim right={(7.8,0)}]  
  
\begin{axis}  
[  
    ybar,  
    enlargelimits=0.15,  
    ylabel={C100 Test Acc.}, 
    ylabel style={at={(-0.08, 0.5)}},
    xlabel={ReLU Budget (\algname{}/ReLU Sharing)},  
    symbolic x coords={50K/78K, 130K/156K, 200K/233K, 300K/311K}, 
    xtick=data,  
    nodes near coords, 
    nodes near coords align={vertical},  
	legend style={at={(0.5,1.11)},
	anchor=north,legend columns=-1},
    ]  
\addplot coordinates {(50K/78K, 67.31) (130K/156K,73.09) (200K/233K, 75.15) (300K/311K, 75.40) };  

\addplot coordinates {(50K/78K,47.24) (130K/156K,60.09) (200K/233K,65.66) (300K/311K, 68.92) };  
\legend{\algname{}, ReLU Sharing}
\end{axis}  
\end{tikzpicture}  } &
        \resizebox{0.45\textwidth}{!}{\begin{tikzpicture}[trim left={(-1.0,0)}, trim right={(7.8,0)}]  
  
\begin{axis}  
[  
    ybar,  
    enlargelimits=0.15,  
    ylabel={C100 Test Acc.}, 
    ylabel style={at={(-0.07, 0.5)}},
    xlabel={ReLU Budget},  
    symbolic x coords={C100-25K, C100-30K, C100-40K}, 
    xtick=data,  
    nodes near coords, 
    nodes near coords align={vertical},  
	legend style={at={(0.5,1.11)},
	anchor=north,legend columns=-1},
    ]  


\addplot+[error bars/.cd, y fixed, y dir=both, y explicit]
        coordinates {   (C100-25K, 66.13) +- (0, 0.09)
                        (C100-30K, 67.37) +- (0, 0.05)
                        (C100-40K, 68.23) +- (0, 0.21)
                        };  

\addplot+[error bars/.cd, y fixed, y dir=both, y explicit] 
        coordinates {   (C100-25K, 65.36) +- (0, 0.16)
                        (C100-30K, 66.77) +- (0, 0.34)
                        (C100-40K, 68.14) +- (0, 0.20)
                        };  


\legend{Vanilla Conv., Separable Conv.}
\end{axis}  
\end{tikzpicture}  } \\
        \multicolumn{1}{c}{\footnotesize (a) \algname{} vs. micro-search space} & 
        \multicolumn{1}{c}{\footnotesize (b) Vanilla Conv vs. Separable Conv} \\
    \end{tabular}
    \vspace{0.9em}
    \caption{Ablation studies: (a) Compares the difference in performance between \algname{} and micro-search space. We leverage the ReLU sharing technique inspired from \cite{NEURIPS2020_c519d47c} on the micro-search space to reduce the ReLU operations without further functionality changes. Furthermore, we remove ReLU layers in the preprocessing layers (which each layer ReLU cost is equivalent to $4{\times}H_i{\times}W_i$). The experiment results show that \algname{} superior to modified micro-search space with above techniques. (b) Compares the operation candidates between vanilla convolution and separable convolution. C100 and Image stand for CIFAR-100, respectively. Vanilla convolution achieves a high test accuracy than separable convolutions given the same network with equivalent \textit{normal}/\textit{reduce} cells except the operation set.}
    \label{fig: ablations studies}
\end{figure}

\subsection{\algname{} vs. Micro-Search Space in ReLU perspective} 
\label{appendix: sphynx vs. micro search space}

\begin{figure}[!ht]
    \centering
    \begin{tabular}{c c}
        \resizebox{0.45\textwidth}{!}{\tikzset{
  annotated cuboid/.pic={
    \tikzset{%
      every edge quotes/.append style={midway, auto},
      /cuboid/.cd,
      #1
    }
    \draw [every edge/.append style={pic actions, densely dashed, opacity=.5}, pic actions]
    (0,0,0) coordinate (o) -- ++(-\cubescale*\cubex,0,0) coordinate (a) -- ++(0,-\cubescale*\cubey,0) coordinate (b) edge coordinate [pos=1] (g) ++(0,0,-\cubescale*\cubez)  -- ++(\cubescale*\cubex,0,0) coordinate (c) -- cycle
    (o) -- ++(0,0,-\cubescale*\cubez) coordinate (d) -- ++(0,-\cubescale*\cubey,0) coordinate (e) edge (g) -- (c) -- cycle
    (o) -- (a) -- ++(0,0,-\cubescale*\cubez) coordinate (f) edge (g) -- (d) -- cycle;
    \path [every edge/.append style={pic actions, |-|}]
    (b) +(0,-5pt) coordinate (b1) edge ["$H_i$ \cubeunits"'] (b1 -| c)
    (b) +(-5pt,0) coordinate (b2) edge ["$H_i$ \cubeunits"] (b2 |- a)
    (c) +(3.5pt,-3.5pt) coordinate (c2) edge ["$4C_i$ \cubeunits"'] ([xshift=3.5pt,yshift=-3.5pt]e)
    ;
  },
  /cuboid/.search also={/tikz},
  /cuboid/.cd,
  width/.store in=\cubex,
  height/.store in=\cubey,
  depth/.store in=\cubez,
  units/.store in=\cubeunits,
  scale/.store in=\cubescale,
  width=10,
  height=10,
  depth=10,
  units=cm,
  scale=.1,
}

\tikzset{
  annotated cuboid2/.pic={
    \tikzset{%
      every edge quotes/.append style={midway, auto},
      /cuboid/.cd,
      #1
    }
    \draw [every edge/.append style={pic actions, densely dashed, opacity=.5}, pic actions]
    (0,0,0) coordinate (o) -- ++(-\cubescale*\cubex,0,0) coordinate (a) -- ++(0,-\cubescale*\cubey,0) coordinate (b) edge coordinate [pos=1] (g) ++(0,0,-\cubescale*\cubez)  -- ++(\cubescale*\cubex,0,0) coordinate (c) -- cycle
    (o) -- ++(0,0,-\cubescale*\cubez) coordinate (d) -- ++(0,-\cubescale*\cubey,0) coordinate (e) edge (g) -- (c) -- cycle
    (o) -- (a) -- ++(0,0,-\cubescale*\cubez) coordinate (f) edge (g) -- (d) -- cycle;
    \path [every edge/.append style={pic actions, |-|}]
    (b) +(0,-5pt) coordinate (b1) edge ["$H_i$ \cubeunits"'] (b1 -| c)
    (c) +(3.5pt,-3.5pt) coordinate (c2) edge ["$C_i$ \cubeunits"'] ([xshift=3.5pt,yshift=-3.5pt]e)
    ;
  },
  /cuboid/.search also={/tikz},
  /cuboid/.cd,
  width/.store in=\cubex,
  height/.store in=\cubey,
  depth/.store in=\cubez,
  units/.store in=\cubeunits,
  scale/.store in=\cubescale,
  width=10,
  height=10,
  depth=10,
  units=cm,
  scale=.1,
}

\tikzstyle{Phase}=[draw, fill=orange!50, text width=2.0em, thick, rotate=90, rounded corners=0.2cm,
    text centered, minimum width=5.0em, minimum height=2.0em, scale=0.6]
\tikzstyle{Phase2}=[draw, fill=green!50, text width=2.0em, thick, rotate=90, rounded corners=0.2cm,
    text centered, minimum width=5.0em, minimum height=2.0em, scale=0.6]
\tikzstyle{Phase3}=[draw, fill=blue!40, text width=2.0em, thick, rotate=90, rounded corners=0.2cm,
    text centered, minimum width=5.0em, minimum height=2.0em, scale=0.6]
\tikzstyle{texto} = [above, text width=20em, text centered, scale=1.0]
\tikzstyle{Cell}=[draw, fill=gray!10, text width=6.0em, thick, rounded corners=0.2cm,
    text centered, minimum width=6.0em, minimum height=3.0em, scale=0.6]
\tikzstyle{sum} = [draw, circle, minimum size=0.1em, thick, scale=0.85]
\newcommand{\preproc}[6]{%
  \begin{pgfonlayer}{background}
    \path (#1.north |- #2.east)+(-0.3, 0.3) node (a1) {};
    \path (#3.south |- #4.west)+(+0.3,-0.3) node (a2) {};
    \path[fill=#5!10,rounded corners, draw=black!50, dashed]
      (a1) rectangle (a2);
    \path (a1.north |- a1.east)+(1.2, -0.34) node (u1)[texto]
      {\tiny #6};
  \end{pgfonlayer}}

\resizebox{1.0\linewidth}{!}{
\begin{tikzpicture}[trim left={(0,0)}, trim right={(6.8,0)}]

  \pic [fill=gray, text=black, draw=black] at (0,0) {annotated cuboid2={width=10, height=10, depth=5, units=}};
  \path (-0.4, 0.2) node (text1) [texto, font=\tiny\linespread{0.8}\selectfont]{\tiny \textbf{Node} (Feature Map)};
  \path (1.5, 0.5) node (relu) [Phase] {ReLU};
  \path (relu.south)+(0.5, 0) node (conv) [Phase2] {Conv};
  \path (conv.south)+(0.5, 0) node (bn) [Phase3] {BN};
  
  \path (1.5, -1.4) node (relu2) [Phase] {ReLU};
  \path (relu2.south)+(0.5, 0) node (conv2) [Phase2] {Conv};
  \path (conv2.south)+(0.5, 0) node (bn2) [Phase3] {BN};
  
  \path (bn2.south) + (1.0, 1.0) node (sum) [sum] {+};
  
  \pic [fill=gray, text=black, draw=black] at (6.0, 0.0) {annotated cuboid2={width=10, height=10, depth=5, units=}};
  \path (5.5, 0.2) node (text2) [texto, font=\tiny\linespread{0.8}\selectfont]{\tiny \textbf{Node} (Feature Map)};
  \path (sum.north)+(0.0, 0.0) node (text3) [texto, font=\tiny\linespread{0.8}\selectfont]{\tiny \textbf{Sum}};
  
  \path [draw, ->, line width=0.4mm] (0.2,-0.28) to (relu.north);
  \path [draw, ->, line width=0.4mm] (relu.south) to (conv.north);
  \path [draw, ->, line width=0.4mm] (conv.south) to (bn.north);
  \path [draw, ->, line width=0.4mm] (bn.south) to (3.9, -0.26);
  
  \path [draw, ->, line width=0.4mm] (0.2,-0.48) to (relu2.north);
  \path [draw, ->, line width=0.4mm] (relu2.south) to (conv2.north);
  \path [draw, ->, line width=0.4mm] (conv2.south) to (bn2.north);
  \path [draw, ->, line width=0.4mm] (bn2.south) to (3.9, -0.49);
  \path [draw, ->, line width=0.4mm] (sum.east) to (5.0, -0.40);
  
  \preproc{relu}{relu}{bn}{bn}{blue}{Convolution Module};
  \preproc{relu2}{relu2}{bn2}{bn2}{blue}{Convolution Module};

\end{tikzpicture}
}} & 
        \resizebox{0.45\textwidth}{!}{\tikzset{
  annotated cuboid/.pic={
    \tikzset{%
      every edge quotes/.append style={midway, auto},
      /cuboid/.cd,
      #1
    }
    \draw [every edge/.append style={pic actions, densely dashed, opacity=.5}, pic actions]
    (0,0,0) coordinate (o) -- ++(-\cubescale*\cubex,0,0) coordinate (a) -- ++(0,-\cubescale*\cubey,0) coordinate (b) edge coordinate [pos=1] (g) ++(0,0,-\cubescale*\cubez)  -- ++(\cubescale*\cubex,0,0) coordinate (c) -- cycle
    (o) -- ++(0,0,-\cubescale*\cubez) coordinate (d) -- ++(0,-\cubescale*\cubey,0) coordinate (e) edge (g) -- (c) -- cycle
    (o) -- (a) -- ++(0,0,-\cubescale*\cubez) coordinate (f) edge (g) -- (d) -- cycle;
    \path [every edge/.append style={pic actions, |-|}]
    (b) +(0,-5pt) coordinate (b1) edge ["$H_i$ \cubeunits"'] (b1 -| c)
    (b) +(-5pt,0) coordinate (b2) edge ["$H_i$ \cubeunits"] (b2 |- a)
    (c) +(3.5pt,-3.5pt) coordinate (c2) edge ["$4C_i$ \cubeunits"'] ([xshift=3.5pt,yshift=-3.5pt]e)
    ;
  },
  /cuboid/.search also={/tikz},
  /cuboid/.cd,
  width/.store in=\cubex,
  height/.store in=\cubey,
  depth/.store in=\cubez,
  units/.store in=\cubeunits,
  scale/.store in=\cubescale,
  width=10,
  height=10,
  depth=10,
  units=cm,
  scale=.1,
}

\tikzset{
  annotated cuboid2/.pic={
    \tikzset{%
      every edge quotes/.append style={midway, auto},
      /cuboid/.cd,
      #1
    }
    \draw [every edge/.append style={pic actions, densely dashed, opacity=.5}, pic actions]
    (0,0,0) coordinate (o) -- ++(-\cubescale*\cubex,0,0) coordinate (a) -- ++(0,-\cubescale*\cubey,0) coordinate (b) edge coordinate [pos=1] (g) ++(0,0,-\cubescale*\cubez)  -- ++(\cubescale*\cubex,0,0) coordinate (c) -- cycle
    (o) -- ++(0,0,-\cubescale*\cubez) coordinate (d) -- ++(0,-\cubescale*\cubey,0) coordinate (e) edge (g) -- (c) -- cycle
    (o) -- (a) -- ++(0,0,-\cubescale*\cubez) coordinate (f) edge (g) -- (d) -- cycle;
    \path [every edge/.append style={pic actions, |-|}]
    (b) +(0,-5pt) coordinate (b1) edge ["$H_i$ \cubeunits"'] (b1 -| c)
    (c) +(3.5pt,-3.5pt) coordinate (c2) edge ["$C_i$ \cubeunits"'] ([xshift=3.5pt,yshift=-3.5pt]e)
    ;
  },
  /cuboid/.search also={/tikz},
  /cuboid/.cd,
  width/.store in=\cubex,
  height/.store in=\cubey,
  depth/.store in=\cubez,
  units/.store in=\cubeunits,
  scale/.store in=\cubescale,
  width=10,
  height=10,
  depth=10,
  units=cm,
  scale=.1,
}

\tikzstyle{Phase}=[draw, fill=orange!50, text width=2.0em, thick, rotate=90, rounded corners=0.2cm,
    text centered, minimum width=5.0em, minimum height=2.0em, scale=0.6]
\tikzstyle{Phase2}=[draw, fill=green!50, text width=2.0em, thick, rotate=90, rounded corners=0.2cm,
    text centered, minimum width=5.0em, minimum height=2.0em, scale=0.6]
\tikzstyle{Phase3}=[draw, fill=blue!40, text width=2.0em, thick, rotate=90, rounded corners=0.2cm,
    text centered, minimum width=5.0em, minimum height=2.0em, scale=0.6]
\tikzstyle{texto} = [above, text width=20em, text centered, scale=1.0]
\tikzstyle{Cell}=[draw, fill=gray!10, text width=6.0em, thick, rounded corners=0.2cm,
    text centered, minimum width=6.0em, minimum height=3.0em, scale=0.6]
\tikzstyle{sum} = [draw, circle, minimum size=0.1em, thick, scale=0.85]
\newcommand{\preproc}[6]{%
  \begin{pgfonlayer}{background}
    \path (#1.north |- #2.east)+(-0.3, 0.3) node (a1) {};
    \path (#3.south |- #4.west)+(+0.3,-0.3) node (a2) {};
    \path[fill=#5!10,rounded corners, draw=black!50, dashed]
      (a1) rectangle (a2);
    \path (a1.north |- a1.east)+(0.8, -0.34) node (u1)[texto]
      {\tiny #6};
  \end{pgfonlayer}}

\resizebox{1.0\linewidth}{!}{
\begin{tikzpicture}[trim left={(-0.5,0)}, trim right={(6.4,0)}]

  \pic [fill=gray, text=black, draw=black] at (0,0) {annotated cuboid2={width=10, height=10, depth=5, units=}};
  \path (-0.4, 0.2) node (text1) [texto, font=\tiny\linespread{0.8}\selectfont]{\tiny \textbf{Node} (Feature Map)};
  \path (1.0, -0.4) node (relu) [Phase] {ReLU};
  \path (relu.south)+(0.8, 1.0) node (conv) [Phase2] {Conv};
  \path (conv.south)+(0.5, 0) node (bn) [Phase3] {BN};
  
  \path (relu.south)+(0.8, -1.0) node (conv2) [Phase2] {Conv};
  \path (conv2.south)+(0.5, 0) node (bn2) [Phase3] {BN};
  
  \path (bn2.south) + (1.0, 1.0) node (sum) [sum] {+};
  
  \pic [fill=gray, text=black, draw=black] at (5.8, 0.0) {annotated cuboid2={width=10, height=10, depth=5, units=}};
  \path (5.5, 0.2) node (text2) [texto, font=\tiny\linespread{0.8}\selectfont]{\tiny \textbf{Node} (Feature Map)};
  \path (sum.north)+(0.0, 0.0) node (text3) [texto, font=\tiny\linespread{0.8}\selectfont]{\tiny \textbf{Sum}};
  
  \path [draw, ->, line width=0.4mm] (0.20,-0.39) to (relu.north);
  \path [draw, ->, line width=0.4mm] (relu.south) to (conv.north);
  \path [draw, ->, line width=0.4mm] (conv.south) to (bn.north);
  \path [draw, ->, line width=0.4mm] (bn.south) to (3.7, -0.26);
  
  \path [draw, ->, line width=0.4mm] (relu.south) to (conv2.north);
  \path [draw, ->, line width=0.4mm] (conv2.south) to (bn2.north);
  \path [draw, ->, line width=0.4mm] (bn2.south) to (3.7, -0.49);
  \path [draw, ->, line width=0.4mm] (sum.east) to (4.8, -0.40);
  
  \preproc{conv}{conv}{bn}{bn}{blue}{Conv. Module};
  \preproc{conv2}{conv2}{bn2}{bn2}{blue}{Conv. Module};

\end{tikzpicture}
}} \\
        \multicolumn{1}{c}{\small (a) Default connection between the Nodes} & 
        \multicolumn{1}{c}{\small (b) Saving ReLU Operations using ReLU Sharing} \\
    \end{tabular}
    \vspace{0.7em}
    \caption{\sl Visualization of saving ReLU counts via \textit{ReLU Sharing}. Unlike \algname{} space, applying \textit{ReLU Sharing} technique to existing NAS search space cannot scale down the network to satisfy low-ReLU budget (e.g., < 70K). (a) Default connection between the nodes in micro-search space. The input feature map needs to go through two ReLU layers for each convolution module. (b) Pre-computing ReLU once and split to each convolution module saves the ReLU count without changing a functionality.}
    \vspace{1em}
    \label{fig: relu sharing}
\end{figure}
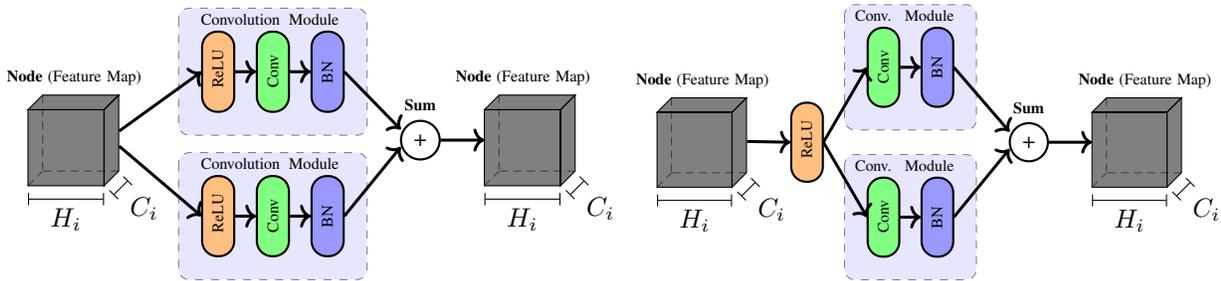

We provide the more comparison between the networks from \algname{} search space and micro-search space evaluating in ReLU perspective. We introduced a ReLU saving technique denoted as ReLU sharing motivated from ReLU shuffling in \cite{NEURIPS2020_c519d47c}. ReLU sharing reduces the number of ReLU operations without changing the functionality. If the input node (or intermediate nodes) connects to two or more convolution modules (ReLU-Conv-BN), ReLU sharing compute the ReLU of intput node once and shares ReLU computed feature map to convolution modules. ReLU sharing reduces ReLU operations more efficiently as more convolution modules attached to the node. However, ReLU sharing acts useless when every node has at most one convolution modules attached in the worst case. Figure~\ref{fig: relu sharing} demonstrates the visual representation of ReLU sharing. If three convolution modules connected to the input node in Figure~\ref{fig: relu sharing} (b), ReLU sharing saves ReLU cost by factor of three. To further reduce the ReLU budget in micro-search space, we remove ReLU layers in the preprocessing layers, 1\texttimes1 convolution modules with ReLU-Conv-BN. We apply ReLU balancing on both methods: ReLU balancing increases channel by factor of 4 when the spatial resolution is halved. 

Figure~\ref{fig: ablations studies} (a) shows additional comparison to \algname{} space and micro-search space with ReLU sharing technique and ReLU pruning on preprocessing layers. The smallest ReLU budget network from the micro-search space has 78K ReLU counts with initial channel $C{=}1$ and $D{=}4$. In case of 154K, 233K, and 311K ReLU networks from micro-search space, we select $C{=}2$, $C{=}3$, and $C{=}4$, respectively while fixing the depth to $D{=}4$. The network skeleton from \algname{} strictly outperforms all range of ReLU budget network to micro-search space based network skeleton, which supports the necessity of \algname{} search space in the lens of ReLU counts.  

\section{Additional ablation studies}
\label{appendix: additional ablation studies}

\subsection{Minuscule deviation in performance in selecting initial channels and depths given a ReLU budget} 
We empirically observe that the network performance deviates minuscule with the possible choice of initial channels and depth to ReLU constraints. We conduct the CIFAR-100 experiments building the networks with the ReLU budget close to 50K with different initial channels (depth changes accordingly). We train five various networks with various initial channels: $C=5,6,7,8,10$. Our experiment excludes the $C=9$ case due to no available setup for a network close to the 50K ReLU budget. We observe that the network performance from various initial channels achieves a small deviation from the mean with a test accuracy of $69.38\pm0.23$ with the best (resp. worst) accuracy of $69.68\%$ (resp. $69.11\%$). Table~\ref{table: channel vs depth} provides a detailed network setup including initial channel, depth, and ReLU counts. 

\begin{table}[ht]
    \centering
    \begin{tabular}{|c | c | c | c | c| c|}
    \hline
    \multirow{2}{*}{\textbf{Method}} & \multicolumn{5}{c|}{\textbf{Init Channel (C) and Depth (D)}} \\
    \cline{2-6} & 
    C=5, D=10 & C=6, D=8 & C=7, D=7 & C=8, D=6 & C=10, D=5 \\
    \hline
    \textbf{ReLU \#} & 51.2K & 49.2K & 50.2K & 49.2K & 51.2K \\
    \hline
    \textbf{Test Accuracy} & 69.64\% & 69.11\% & 69.68\% & 69.27\% & 69.22\% \\
    \hline
    \end{tabular}
    \vspace{0.8em}
    \caption{\sl Detailed experiment setup for selecting channels and depth. Given various initial channels $C=5,6,7,8, 10$, we choose the depth such that the networks' ReLU count close to 50K as possible. In this case, the ReLU count formula is $32{\times}32{\times}C{\times}D$ due to ReLU balancing technique. $32$ comes from the original image size of CIFAR-100. We observe that the network performs consistently on various C and D selections with small standard deviations. ($69.38\pm 0.23$)}
    \vspace{-0.15em}
    \label{table: channel vs depth}
\end{table}

\subsection{ReLU Balancing vs. FLOP Balancing} We compare the performance between ReLU balancing and FLOP balancing. As a recall, FLOP balancing is the typical rule that many existing convolutional neural networks follow, which increase channel size by a factor of two when the spatial resolution reduces by half. We compare four different ReLU budget networks: 25K, 30K, 40K, and 50K. First, we leverage the same \textit{normal}/\textit{reduce} cells in Figure~\ref{appendix: cell pictures} to construct the network. Second, we match the equivalent depth for the FLOP balancing network and reduce cells' locations to the ReLU balancing network. Finally, we adjust initial channels on the FLOP balancing network to satisfy a given ReLU budget. As shown in Table~\ref{table: relu vs flop} ReLU balancing applied networks outperforms the FLOP balancing networks in test accuracy. Furthermore, FLOP balanced networks struggle with inconsistency in the performance of the given ReLU budget. Table~\ref{table: relu vs flop experiment setup} includes the network configurations conducted in this experiment. Table~\ref{table: relu vs flop experiment setup} shows the experiment setup, including the network initial channels, depth, and reduce cell locations. We note that we carefully chose $C$ for a ReLU budget while fixing $D$ such that reducing cell cases takes the same advantage of reducing cell locations.  

\begin{table}[!ht]
    \centering
    \vspace{-1.0em}
    \begin{tabular}{| c | c | c | c | c|}
    \hline
    \multirow{2}{*}{\textbf{Method}} & \multicolumn{4}{c|}{\textbf{ReLU Budget} (ReLU Bal. Net. / FLOP Bal. Net.)} \\
    \cline{2-5} & 25.6K/26.1K & 30.2K/30.4K & 41.0K/41.4K & 51.2K/53.7K \\
    \hline
    ReLU Bal. Test Acc. & \textbf{66.16}\% & \textbf{67.31}\% & \textbf{68.39}\% & \textbf{69.64}\% \\ 
    \hline
    FLOP Bal. Test Acc. & 65.14\% & 66.16\% & 65.55\% & 69.32\% \\
    \hline
    \end{tabular}
    \vspace{0.8em}
    \caption{\sl Comparison between the network based on ReLU balancing and FLOP balancing on CIFAR-100. Bal. stands for balancing. Applying ReLU balancing on a cell-based network achieves a higher test accuracy over FLOP balancing. Furthermore, ReLU balancing enjoys consistency on test accuracy improvement given the ReLU budget, while FLOP balancing does not. (e.g., FLOP balancing network on 41.4K ReLU budget)} 
    \vspace{-0.15em}
    \label{table: relu vs flop}
\end{table}

\begin{table}[ht]
    \centering
    \begin{tabular}{|c | c | c | c | c|}
    \hline
    \multirow{2}{*}{\textbf{Method}} & \multicolumn{4}{c|}{\textbf{ReLU Budgets} (ReLU Balancing/FLOP Balancing)} \\
    \cline{2-5} & 
    25.6K/26.1K & 30.2K/30.4K & 41.0K/41.4K & 51.2K/53.7K \\
    \hline
    ReLU Balancing & C=5, D=5 & C=5, D=6 & C=5, D=8 & C=5, D=10 \\
    FLOP Balancing & C=17, D=5 & C=17, D=6 & C=12, D=8 & C=14, D=10\\
    \hline
    Reduce Cell Loc. & (0, 1) & (0, 1) & (1, 3) & (0, 5) \\
    \hline
    \end{tabular}
    \vspace{0.8em}
    \caption{\sl Experiment setup for ReLU balancing and FLOP balancing comparison. We use equivalent cell structures in Figure~\ref{fig: cells} to construct the network. We use the same reduce cells location for both network. Note that the number of ReLU operations in FLOP balancing affects by \textit{reduce} cells locations. }
    \vspace{-0.15em}
    \label{table: relu vs flop experiment setup}
\end{table}


\subsection{Vanilla Convolution vs. Separable Convolution} This section compares the performance difference between vanilla convolution (which we adopt in our search space) and separable depthwise convolution. We leverage the same cells from our methods and convert the vanilla convolutions with separable convolutions without ReLU layers. Our empirical results on CIFAR-100 in Figure~\ref{fig: ablations studies} (b) show that vanilla convolution achieves a slightly higher test accuracy compare to the separable convolutions. Further, we test on a small network with 188K ReLU counts on ImageNet, in which we observe the more significant performance difference as 38.11\% and 35.95\% with vanilla and separable convolution, respectively. Table~\ref{table: vanilla vs separable details} includes the details in formations, including test performance and model parameters. We observe that separable convolution-based networks struggle with small network parameter size, especially in the small-ReLU budget network. 

\begin{table}[ht]
    \centering
    \begin{tabular}{|c | c | c | c |}
    \hline
    \multirow{2}{*}{\textbf{Method}} & \multicolumn{3}{c|}{\textbf{ReLU Budgets}} \\
    \cline{2-4} & 
    25.6K & 30.2K & 41.0K  \\
    \hline
    Vanilla Conv. & $66.13\pm0.09$\% & $67.37\pm0.05$\% & $68.23\pm0.21$\%  \\
    Separable Conv. & $65.36\pm0.16$\% & $66.77\pm0.34$\% & $68.14\pm0.20$\% \\
    \hline
    Param. (Vanilla/Separable). & 2.75M/0.35M & 3.41M/0.43M & 3.45M/0.45M \\
    \hline
    \end{tabular}
    \vspace{0.8em}
    \caption{\sl Comparison between the network with vanilla convolutions and separable convolutions.}
    \vspace{-0.15em}
    \label{table: vanilla vs separable details}
\end{table}

\subsection{\textit{Normal} and \textit{Reduce} cells with smaller number of nodes}
We recall our cell structures. We define the operation set, eliminating the ReLU layer in convolution modules into a sequence of Conv-BN from ReLU-Conv-BN. We also remove the ReLU layers in 1\texttimes1 preprocessing layers. In other words, only linear operations exist inside the cells. We ask whether to reduce the linear operations since the operations inside the cells are equivalent to linear combinations of linear functions. 

\begin{figure}[!ht]
    \centering
    \renewcommand{\arraystretch}{0.2}
    \def\sw{0.45\linewidth}
    \begin{tabular}{c c}
        \includegraphics[width=\sw]{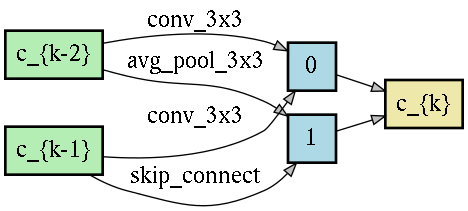} &
        \includegraphics[width=\sw]{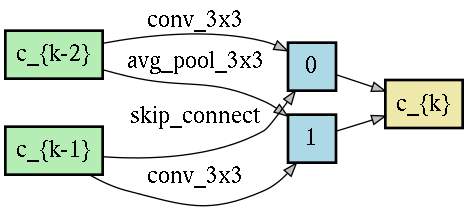} \\
        \algname{}-$N=5$ Normal Cell & \algname{}-$N=5$ Reduce Cell \\
    \end{tabular}
    \vspace{1.5em}
    \caption{Smaller \textit{Normal} and \textit{reduce} cells found by \algname{} with $N=5$ from CIFAR-100.}
    \vspace{1.5em}
    \label{fig: small cells}
\end{figure}

We conduct the experiments by leveraging DARTs designing the cells $N=5$. We first search the cells with DARTs with identical training protocols to Section. Figure~\ref{fig: small cells} shows the cells we found with $N=5$ nodes. Then, conducting apple-to-apple comparisons between $N=7$ nodes and $N=5$ nodes with identical training hyperparameters, we observe the superior performance in test accuracy with $N=7$ nodes cells over $N=5$ nodes. 

\begin{table}[ht]
    \centering
    \begin{tabular}{|c | c | c | c | c|}
    \hline
    \multirow{2}{*}{\textbf{Method}} & \multicolumn{4}{c|}{\textbf{ReLU Budgets}} \\
    \cline{2-5} & 25K & 30K & 40K & 50K \\
    \hline
    \algname{} $N=5$ Cells & 64.24\% & 66.51\% & 67.92\% & 68.80\% \\
    \hline
    \algname{} $N=7$ Cells& $\bm{66.13\pm0.09}$\% & $\bm{67.37\pm0.05}$\% & $\bm{68.23\pm0.21}$\% &  $\bm{65.57\pm0.06}$\% \\ 
    \hline
    \end{tabular}
    \vspace{0.8em}
    \caption{\sl Test accuracy comparison between $N=5$ cells (smaller cells) and $N=7$ cells (original) found from \algname{}.}
    \vspace{-0.15em}
    \label{table: small cells comparison}
\end{table}

\subsection{Recycling DARTS/PC-DARTS cells to \algname{} Search Space}
What will happen if we skip the search phase and evaluate the final network applying existing NAS cells such as DARTS/PC-DARTS cells? We modify the cells following the rule in Section~\ref{subsec: Efficient search space}:
\begin{enumerate}
    \itemsep-0.10em 
    \item We replace separable convolution with regular convolution.
    \item We change the convolution module from ReLU-Conv-BN to Conv-BN.
    \item We replace max-pooling layers with average pooling layers.
    \item We add the non-linear layers at the end of each cell.
\end{enumerate}
These modification, compare to Appendix~\ref{appendix: sphynx vs. micro search space}, allows designing ReLU-efficient networks with existing NAS cells. Our ablation study shows that the cells found directly from the \algname{} perform superior to modified DARTS and PC-DARTS cells searched from the original micro-search space as shown in Table~\ref{table: darts cells comparison}.   


\begin{table}[ht]
    \centering
    \begin{tabular}{|c | c | c | c | c|}
    \hline
    \multirow{2}{*}{\textbf{Method}} & \multicolumn{4}{c|}{\textbf{ReLU Budgets}} \\
    \cline{2-5} & 25K & 30K & 40K & 50K \\
    \hline
    DARTS Cell & 59.86\% & 62.03\% & 66.92\% & 68.60\% \\
    PC-DARTS Cell & 64.32\% & 66.24\% & 67.78\% & 69.07\%\\
    \hline
    \algname{} & $\bm{66.13\pm0.09}$\% & $\bm{67.37\pm0.05}$\% & $\bm{68.23\pm0.21}$\% &  $\bm{65.57\pm0.06}$\% \\ 
    \hline
    \end{tabular}
    \vspace{0.8em}
    \caption{\sl The final architecture performance (equivalent reduce locations) comparison between the cells \algname{} found to existing cells from DARTS and PC-DARTS. Separable convolutions, separable dilated convolution, and max-pooling replaced by vanilla convolution, vanilla dilated convolution, and average pooling, respectively.}
    \vspace{-0.15em}
    \label{table: darts cells comparison}
\end{table}

\begin{figure}[!ht]
    \centering
    \renewcommand{\arraystretch}{0.2}
    \def\sw{0.40\linewidth}
    \begin{tabular}{c c}
        \includegraphics[width=\sw]{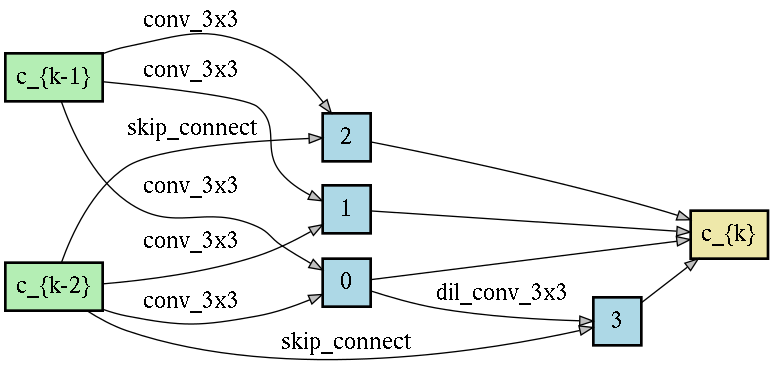} &
        \includegraphics[width=\sw]{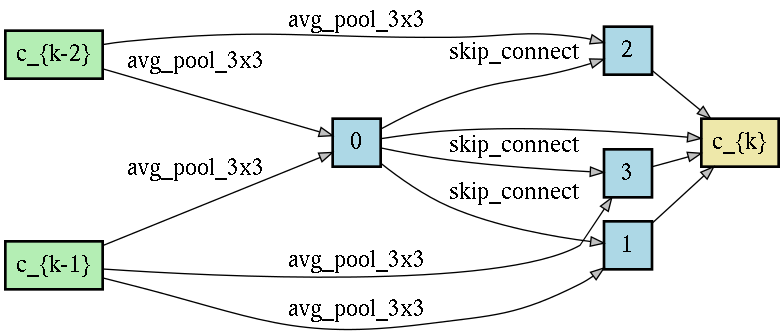} \\ 
        DARTS \textit{Normal} Cell & DARTS \textit{Reduce} Cell \\
        \vspace{1.5em}
        \includegraphics[width=\sw]{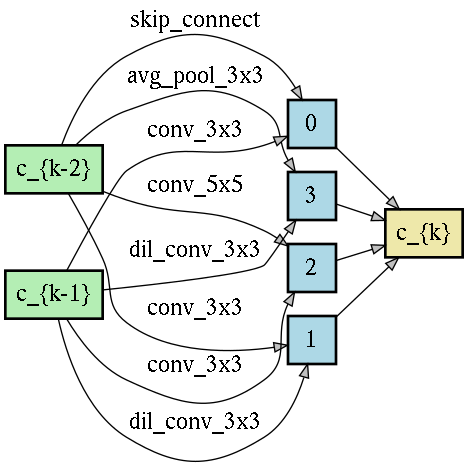} &
        \includegraphics[width=\sw]{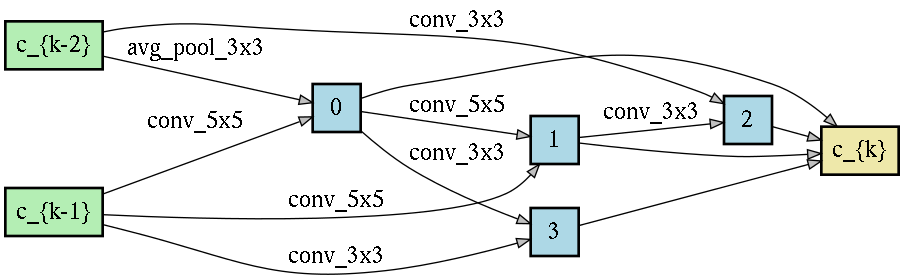} \\
        PC-DARTS \textit{Normal} Cell & PC-DARTS \textit{Reduce} Cell \\
    \end{tabular}
    \vspace{1.5em}
    \caption{Converted DARTS and PC-DARTS \textit{normal}/\textit{reduce} cells for \algname{} design space.}
    \vspace{1.5em}
    \label{fig: darts and pcdarts cells}
\end{figure}

\section{Boosting Performance with Knowledge Distillation}
\label{appendix: knowledge distillation}
This section includes a direct comparison to the DeepReDuce paper on the CIFAR-100 dataset with knowledge distillation. DeepReDuce empirically showed that leveraging KD with the pretrained teacher model with original network architecture and ReLU pruned student network boosts the test accuracy over regular training. 
We apply the KD methods from \cite{NIPS2014_ea8fcd92}, which reformulates the objective function by adding the regression mean square error loss between teacher's and student's logit. Finally, we formally state the objective functions for the knowledge distillation. 

Let $f_s$ and $f_t$ be the student and teacher network, respectively. Let $f_s$ is parameterized by $\rvw$. Let $g_s$ and $g_t$ be the network logit corresponding from $f_s$ and $f_t$, respectively. Given a train (T) and validation (V) dataset of labeled pairs (x, y) drawn from a joint distribution (X, Y). The KD objective functions is defined as following:
\begin{equation}
    \min_{\rvw \in \R^d} \sum_{(x, y) \in T} \mathcal{L}(f_s(\rvw, x), y) + \|g_s(\rvw, x) - g_t(x)\|_2^2
\end{equation}

This section includes a direct comparison to the DeepReDuce paper on the CIFAR-100 dataset with knowledge distillation. DeepReDuce empirically showed that leveraging KD with the pretrained teacher model with original network architecture and ReLU pruned student network boosts the test accuracy over regular training. 
We apply the KD methods from \cite{NIPS2014_ea8fcd92}, which reformulates the objective function by adding the regression mean square error loss between teacher's and student's logit. Finally, we formally state the objective functions for the knowledge distillation. 

Our teacher model achieves $76.20\%$ test accuracy, and the network consists of equivalent cells we found from \algname{} but replacing the convolution modules from Conv-BN to ReLU-Conv-BN. Table~\ref{table: CIFAR100 KD} shows the direct comparison of \algname{} and DeepReDuce. Our results show that \algname{} overall achieves the better test accuracy for the ReLU budget except for the DeepReDuce network with a 24.6K ReLU budget. In the PI latency perspective, DeepReDuce with 24.6K ReLU network outperforms \algname{} 25.6K ReLU network by 117ms faster. While the online latency consists of computing the linear layer and non-linear layer from GC, the linear layer computation time may not be neglectable in an extremely low-ReLU budget network. We conjecture that the linear layer computation in \algname{} is taking more than DeepReDuce due to the network size resulting in significant PI latency although small ReLU budget difference (24.6K vs. 25.6K).

\begin{table}[ht]
\centering
\caption{Knowledge Distillation Results on CIFAR-100. The lower the ReLUs and PI latency the better.}
\begin{threeparttable}
    \begin{tabular}{c c c | c c c | c c c}
        \toprule
        \multicolumn{3}{c}{\textbf{DeepReDuce}} & \multicolumn{3}{c}{\textbf{\algname{}}} & \multicolumn{3}{c}{\textbf{Improvement}} \\
        \midrule
        \multicolumn{1}{c}{ReLUs} & \multicolumn{1}{c}{Test Acc.} & \multicolumn{1}{c|}{PI Lat.} & \multicolumn{1}{c}{ReLUs} & \multicolumn{1}{c}{Test Acc.} & \multicolumn{1}{c|}{PI Lat.} &
        \multicolumn{1}{c}{ReLUs} & \multicolumn{1}{c}{Test Acc.} & \multicolumn{1}{c}{PI Lat.} \\ 
        \midrule
        \multicolumn{1}{c}{24.6K} & \multicolumn{1}{c}{68.41\%} & \multicolumn{1}{c|}{579ms} & \multicolumn{1}{c}{25.6K} & \multicolumn{1}{c}{68.40\%} & \multicolumn{1}{c|}{727ms} & \multicolumn{1}{c}{1.04\texttimes} & \multicolumn{1}{c}{+0.16} & \multicolumn{1}{c}{+117ms} \\
        \multicolumn{1}{c}{28.7K} & \multicolumn{1}{c}{68.70\%} & \multicolumn{1}{c|}{738ms} & \multicolumn{1}{c}{31.2K} & \multicolumn{1}{c}{70.21\%} & \multicolumn{1}{c|}{861ms} & \multicolumn{1}{c}{1.08\texttimes} & \multicolumn{1}{c}{+1.51} & \multicolumn{1}{c}{+123ms} \\
        \multicolumn{1}{c}{49.2K} & \multicolumn{1}{c}{71.10\%} & \multicolumn{1}{c|}{1190ms} & \multicolumn{1}{c}{41.0K} & \multicolumn{1}{c}{71.57\%} & \multicolumn{1}{c|}{1114ms} & \multicolumn{1}{c}{0.83\texttimes} & \multicolumn{1}{c}{+0.47} & \multicolumn{1}{c}{-76ms} \\
        \multicolumn{1}{c}{57.3K} & \multicolumn{1}{c}{72.68\%} & \multicolumn{1}{c|}{1350ms} & \multicolumn{1}{c}{51.2K} & \multicolumn{1}{c}{72.85\%} & \multicolumn{1}{c|}{1335ms} & \multicolumn{1}{c}{0.89\texttimes} & \multicolumn{1}{c}{+0.17} & \multicolumn{1}{c}{-15ms} \\
        \bottomrule
    \end{tabular}
\end{threeparttable}
\label{table: CIFAR100 KD}
\end{table}

\section{Architecture Details for ImageNet}
\label{appendix: architecture details for imagenet}

We transfer the cells we learned from CIFAR-100 to a complex image classification dataset. Primarily, we construct a slightly modified network following from DARTS. We start the network with three convolution stem layers with a stride of 2, reducing the input image resolution from 224\texttimes 224 to 28 \texttimes 28. We note that the first stem convolution is Conv-BN's sequence and increases the channel dimension from 3 to $C/2$. The second and third stem convolution module has ReLU-Conv-BN sequence. The second stem convolution increase the channel dimensions from $C/2$ and $C$, and the third stem convolution maintains the channel dimensions but only reduces the feature map size by half. Then the cells are stacked, taking second and third stem convolutions as inputs.

\section{Details for Final Networks from \algname{} for Different ReLU Budgets}
\label{appendix: network config for ReLU budgets}

As described in Section~\ref{subsec: searching cells to sphynx space}, the number of initial channels and network depth determine the total ReLU count. We provide these two hyperparameters we used for various ReLU budgets on CIFAR-100, Tiny-ImageNet, and ImageNet in Table~\ref{table: network configs}. Also, we offer the VGG11 and ResNet18 network specifications for various ReLU budgets we used as the comparison. For both VGG11 and ResNet18, we only control the initial number of channels to satisfy the desired ReLU budget.

\begin{table}[!ht]
    \centering
    \caption{Detailed network configurations at various ReLU budgets for \algname{}, VGG11, and ResNet18.}
    \begin{threeparttable}
        \begin{tabular}{c|c c c c}
            \toprule
            \multicolumn{1}{c|}{\textbf{Dataset}} & \multicolumn{1}{c}{\textbf{Init. Ch.}} & \multicolumn{1}{c}{\textbf{Network Depth}} & \multicolumn{1}{c}{\textbf{ReLUs}} & \multicolumn{1}{c}{\textbf{Reduce Cells Loc.}} \\
            \midrule
            CIFAR-100 & 5 & 5 & 25.6K & $(0, 1)$    \\
            CIFAR-100 & 5 & 6 & 30.2K & $(0, 1)$    \\
            CIFAR-100 & 5 & 8 & 41.0K & $(1, 3)$    \\
            CIFAR-100 & 5 & 10 & 51.2K & $(0, 5)$   \\
            CIFAR-100 & 7 & 10 & 71.7K & $(0, 5)$   \\
            CIFAR-100 & 10 & 10 & 102.4K & $(0, 5)$ \\
            CIFAR-100 & 15 & 15 & 230.0K & $(2, 6)$ \\
            \midrule
            Tiny-ImageNet & 5 & 5 & 102.4K & $(0, 1)$    \\
            Tiny-ImageNet & 5 & 10 & 204.8K & $(0, 5)$    \\
            Tiny-ImageNet & 7 & 10 & 286.7K & $(0, 5)$    \\
            Tiny-ImageNet & 20 & 10 & 819.2K & $(0, 5)$    \\
            \midrule
            ImageNet & 10 & 10 & 172K & $(1, 5)$\\
            ImageNet & 20 & 10 & 345K & $(1, 5)$\\
            ImageNet & 30 & 10 & 517K & $(1, 5)$\\
            ImageNet & 40 & 10 & 690K & $(1, 5)$\\ 
            ImageNet & 50 & 10 & 862K & $(1, 5)$\\
            ImageNet & 60 & 10 & 1034K & $(1, 5)$\\
            \midrule
            ImageNet (VGG11) & 4 & 11 & 472K & N/A \\
            ImageNet (VGG11) & 8 & 11 & 936K & N/A \\
            ImageNet (VGG11) & 64 & 11 & 7488K & N/A \\
            \midrule
            ImageNet (ResNet18) & 8 & 18 & 288K & N/A \\
            ImageNet (ResNet18) & 16 & 18 & 577K & N/A \\
            ImageNet (ResNet18) & 32 & 18 & 1154K & N/A \\
            ImageNet (ResNet18) & 64 & 18 & 2308K & N/A \\
            \bottomrule
        \end{tabular}
    \end{threeparttable}
    \label{table: network configs}
\end{table}

\end{document}